\documentclass[10pt]{amsart}
\usepackage{latexsym,amssymb,amsmath,
}

\newcommand{\arr}{\longrightarrow}

\newcommand{\mx}[1]{\mbox{{#1}}}
\newcommand{\s}{\quad}

\newcommand{\Sym}{{\tt S}}

\DeclareMathOperator{\Ind}{Ind}
\DeclareMathOperator{\Sing}{Sing}
\def\wt{{\mathrm{wt}\,}}

\newcommand{\dst}{\mbox{{\sf d}}}
\newcommand{\dvar}[1]{\mbox{{\sf d}$x_{#1}$}}
\newcommand{\ddind}[2]{\mbox{{\sf d}$_{#1#2}$}}
\newcommand{\dudind}[2]{\mbox{{\sf d}$^{#1}_{#2}$}}
\newcommand{\dprt}{\mbox{$\partial$}}
\newcommand{\dpind}[1]{\mbox{${\partial}_{#1}$}}
\newcommand{\dhind}[1]{\mbox{$\hat{\partial}_{#1}$}}

\newcommand{\lsp}[1]{\langle {#1} \rangle}
\newcommand{\pmo}{\mx{$\scriptstyle{(+-)}$}}
\newcommand{\mpo}{\mx{$\scriptstyle{(-+)}$}}
\newcommand{\lag}{{\tt g}}
\newcommand{\gind}[1]{{\tt g}_{#1}}

\newcommand{\DIV}{\mathop{\rm div\,}}
\newcommand{\al}{\alpha}

\newcommand{\de}{\delta}
\newcommand{\De}{\Delta}
\newcommand{\ep}{\varepsilon}

\newcommand{\La}{\Lambda}

\newcommand{\si}{\sigma}

\newcommand{\Om}{\Omega}

\def\Bw{{\mathcal B}}

\def\Dw{{\mathcal D}}

\def\Hw{{\mathcal H}}

\def\Mw{{\mathcal M}}
\def\Nw{{\mathcal N}}

\def\Pw{{\mathcal P}}

\def\CC{{\mathbb{C}}}

\def\PP{{\mathbb{P}}}

\def\ZZ{{\mathbb{Z}}}

\newcommand{\fg}{\mbox{{\tt g}}}

\newcommand{\ad}{\mathop{\rm ad \, }}
\renewcommand{\Re}{\mathop{\rm Re}}

\renewcommand{\tilde}{\widetilde}
\renewcommand{\theequation}%
  {\arabic{section}.\arabic{equation}}
\makeatletter
\renewcommand\section%
 {\@startsection {section}{1}{\z@}%
  {-3.5ex \@plus -1ex \@minus -.2ex}%
    {2.3ex \@plus.2ex}%
       {\normalfont\large\bfseries}}
\@addtoreset{equation}{section}
\makeatother

\newtheorem{Proposition}{Proposition}[section]
\newcommand{\bPr}{\begin{Proposition}}
\newcommand{\ePr}{\end{Proposition}}

\newtheorem{Theorem}[Proposition]{Theorem}
\newcommand{\bTh}{\begin{Theorem}}
\newcommand{\eTh}{\end{Theorem}}

\newtheorem{Lemma}[Proposition]{Lemma}
\newcommand{\bLe}{\begin{Lemma}}
\newcommand{\eLe}{\end{Lemma}}

\newtheorem{Definition}[Proposition]{Definition}
\newcommand{\bDe}{\begin{Definition}}
\newcommand{\eDe}{\end{Definition}}

\newtheorem{Corollary}[Proposition]{Corollary}
\newcommand{\bCo}{\begin{Corollary}}
\newcommand{\eCo}{\end{Corollary}}

\newtheorem{Conjecture}[Proposition]{Conjecture}
\newcommand{\bCj}{\begin{Conjecture}}
\newcommand{\eCj}{\end{Conjecture}}

\newtheorem{remark}[Proposition]{Remark}

\newcommand{\bEq}{\begin{equation}}
\newcommand{\eEq}{\end{equation}}

\newcommand{\bEa}{\begin{eqnarray}}
\newcommand{\eEa}{\end{eqnarray}}

\newcommand{\bEaz}{\begin{eqnarray*}}
\newcommand{\eEaz}{\end{eqnarray*}}

\newcommand{\bAr}{\begin{array}}
\newcommand{\eAr}{\end{array}}

\newcommand{\bN}{\begin{enumerate}}
\newcommand{\eN}{\end{enumerate}}

\newcommand{\bD}{\begin{description}}
\newcommand{\eD}{\end{description}}

\newcommand{\prf}{{\sl Proof.}}

\newcommand{\epf}{$\Box$}

\newcommand{\alphaparenlist}{%
  \renewcommand{\theenumi}{\alph{enumi}}%
  \renewcommand{\labelenumi}{(\theenumi)}%
}

\newcommand{\arabicparenlist}{%
  \renewcommand{\theenumi}{\arabic{enumi}}%
  \renewcommand{\labelenumi}{(\theenumi)}%
}
\newcommand{\romanparenlist}{%
  \renewcommand{\theenumi}{\roman{enumi}}%
  \renewcommand{\labelenumi}{(\theenumi)}%
}
\newcommand{\starlist}{%
  \renewcommand{\theenumi}{\rm\textup{(*)}}%
  \renewcommand{\labelenumi}{\theenumi}%
}

\begin{document}

\title[Representations of $E(3,6)$.I: Irreducible modules.]
{Representations of the exceptional
Lie superalgebra
$E(3,6)$: I.~Degeneracy conditions.
}

\author{Victor G. Kac${}^*$ and Alexei Rudakov${}^\dagger$}
\thanks{${}^*$~Supported in part by NSF grant
    DMS-9970007. \\
\hspace*{1.5em}${}^\dagger$~Research
partially financed by {\it European Commission}
(TMR 1998-2001 Network {\it Harmonic Analysis}).
}

\begin{abstract}
Recently one of the authors obtained a classification of simple
infinite-dimensional Lie superalgebras of vector fields which
extends the well-known classification of E.~Cartan in the Lie
algebra case.  The list consists of many series defined by simple
equations, and of several exceptional superalgebras, among them $E(3,6)$.

In the article we study irreducible representations
of the exceptional Lie superalgebra $E(3,6)$. This superalgebra has
$s\ell(3)\times s\ell(2)\times g\ell(1)$ as the zero degree component
of its consistent $\ZZ$-grading which leads us to believe that its
representation theory has potential
for physical applications.
\end{abstract}
\maketitle

\section{Introduction.}

Recently V.~Kac obtained the classification of infinite-dimensional
simple linearly compact Lie superalgebras [K1]. Two of the
exceptional algebras, $E(3,6)$ and $E(3,8)$ have the Lie algebra
$s\ell(3)\times s\ell(2)\times g\ell(1)$ as the zero degree component
$\gind 0$ in their
consistent $\ZZ$-grading. This points to the potential physical applications
of representations of these algebras.

We deal with representations of $E(3,6)$
in this article having the main objective to classify and describe
irreducible representations.

We follow the approach developed for representations of infinite-dimensional
simple linearly compact Lie algebras by A.~Rudakov in [R].
The problem reduces
quite quickly (Proposition~\ref{prop:1.3}) to the description of
the so called
degenerate modules, and for the latter we have to study singular
vectors and secondary singular vectors (see definitions in Section~1).

In this first article on the topic we get
an important restriction on the list of degenerate
irreducible representations of $E(3,6)$ (Theorem~\ref{T:sl3-weights}).

In order to obtain the complete list of these
representations and to get a hold on
their construction and  structure more work is to be done.  We describe it
in the subsequent articles.  In particular, it turns out that the
degenerate irreducible representations of $E (3,6)$ fall into
four series, and we construct four complexes of $E(3,6)$-modules
which lead to an explicit description of these series.

Let us mention that the existence of exceptional superalgebras  was
announced by I.~Shche\-poch\-ki\-na [S1] in 1983,
but her construction is implicit and
quite difficult to use (see~[S2]).
We rely on the explicit construction
of $E(3,6)$ found  by S.J.~Cheng and V.~Kac ([K1, CK1]).

For the related mathematical development see [K2, K3, CK2].
Basic properties of superalgebras can be found for example in [K4].

All vector spaces, linear maps and tensor products are considered
over $\CC$.

We would like to thank IHES, where this work was started in the
spring of 1999, and the Schr\"odinger Institute, where this work
has been completed, for their hospitality.  One of the authors
wishes to thank E.B.~Vinberg for valuable correspondence.

\section{General remarks on representations of linearly compact Lie
superalgebras.}

Representation theory of infinite-dimensional Lie superalgebras
was initiated by A~Rudakov some 25~years ago. We will follow the
same approach in the Lie superalgebra case. It is a worthwhile undertaking
because the list of linearly compact infinite-dimensional simple Lie
superalgebras and their irreducible modules turns out to be much richer than
in the Lie algebra case and because of their potential applications to
quantum physics.

It is most natural to consider continuous representations of linearly
compact Lie algebras in linearly compact topological spaces.
However, technically it is more convenient to work with the contragredient
to these, which are continuous representations in spaces with discrete
topology. The continuity of a representation of a linearly
compact Lie superalgebra $L$ in a vector space $V$ with discrete
topology means that the stabilizer
$L_v=\{ g \in L \, | \,gv=0 \}$
of any $v\in V$ is an open (hence of finite codimension) subalgebra of
$L$.

Let $L_0$ be an open subalgebra of $L$. In order to avoid pathological
examples we shall always assume that the $L$-module $V$ is $L_0$-locally
finite, meaning that any $v \in V$ is contained in a finite-dimensional
$L_0$-invariant subspace. We shall denote by $\Pw (L,L_0)$ the category
of all continuous $L$-modules $V$, where $V$ is a vector space with discrete
topology, that are $L_0$-locally finite.
When talking about representations of $L$, we shall always mean modules from
$\Pw(L,L_0)$ (after a suitable choice of $L_0$), unless otherwise is stated.

Suppose that $L$ is a simple infinite-dimensional linearly compact Lie
superalgebra. It is known [K1] that $L$ has a maximal open subalgebra $L_0$
called a
\textit{maximally even}  subalgebra
which contains all even exponentiable elements of $L$
(in most of the cases $L$ has
a unique such subalgebra).
If $L'_0$ is another open subalgebra of $L$,
then, since all even elements of $L'_0$ are exponentiable [K1], we conclude
that the even part of $L'_0$ lies in $L_0$.  It follows that
$$ \Pw(L,L'_0) \supset  \Pw(L,L_0), $$
and therefore any open subalgebra of $L$ acts locally finitely
on modules from $\Pw(L,L_0)$.

Let $L_-$ be a complementary to $L_0$ (finite-dimensional) subspace in
$L$. In most (but not all cases) of simple $L$ and maximally even $L_0$
one can choose $L_-$ to be a subalgebra. Choosing an ordered basis of
$L_-$ we denote by $U(L_-)$ the span of all PBW monomials in this basis.
We have:  $U(L)=U(L_-)\otimes\,U(L_0)$, a vector space tensor product.
(Here and further $U(L)$ stands for the universal enveloping algebra of
the Lie algebra $L$.)
It follows that any irreducible $L$-module $V$ from the category
$\Pw(L,L_0)$ is finitely generated over $U(L_-)$:
$$ V=U(L_-)E $$
for some finite-dimensional subspace $E$. This last property is very
important in the theory of conformal modules [CK2], [K2].\\

Let $V$ be an $L$-module from the category $\Pw(L,L_0)$. Denote by
$\Sing V$ the sum of all irreducible $L_0$-submodules of $V$. This
subspace is clearly different from zero. Its vectors are called
\textit{singular vectors} of the $L$-module $V$.


Given an $L_0$-module $F$, we may consider the associated induced
$L$-module
\begin{displaymath}
  M(F) =\Ind_{L_0}^L F =  U(L) \otimes_{U(L_0)}F \, ,
\end{displaymath}
called also the \emph{universal} $L$-module (associated to $F$).
Other names used for these kinds of modules are:  generalized
Verma modules, Weyl modules, etc.

The $L_0$-module $F$ is canonically an $L_0$-submodule of $M(F)$,
and the sum of its irreducible submodules, that is
$\Sing F$, is a subspace of $\Sing M(F)$, called \textit{the subspace of
trivial singular vectors.}

Let us mention that if $F$ is finite-dimensional then
being  a continuous $L_0$-module
it is annihilated by an open ideal $I$ of $L_0$,
so, in fact, in this case $F$ is a module
over a finite-dimensional Lie superalgebra $L_0/I$.

The following proposition is standard.

\bPr
  \label{prop:1.1}
\alphaparenlist
\begin{enumerate}
\item 
  A finite-dimensional $L_0$-module $F$ is continuous if and only if
$Ann F=\{ g \in L_0 |\,\, gF=0 \}$ is an open ideal of $L_0$.

\item  
  If $L$ has a filtration by open subalgebras:  $L = L_{-1}
  \supset L_0 \supset L_1 \supset \cdots$ and $F$ is a continuous
  finite-dimensional $L_0$-module, then the $L$-module $M(F)$
  lies in $\Pw(L,L_0)$.

%

\end{enumerate}
\ePr
\prf\/
(a) is trivial.  Hence, if $F$ is a continuous finite-dimensional
  $L_0$-module, we have:  $L_j F=0$ for $j \gg 0$.  Note that $
 M(F)=U(L_-)F$, hence we can make an increasing filtration of $M(F)$ by
 finite-dimensional subspaces:
 \begin{displaymath}
   F \subset L_-F+F \subset L^2_- F+L_-F+F \subset \cdots \, .
 \end{displaymath}
But, clearly, each member of this filtration is annihilated by
$L_j$ for $j \gg 0$, which proves the continuity of the
$L$-module $M(F)$.  Since $\dim L_- < \infty$, we conclude also
that $M(F)$ is $L_0$-locally finite, proving~(b).
\epf

\bDe An irreducible $L$-module $V$ is called non-degenerate if
$V=M(F)$ for  an irreducible $L_0$-module $F$.
We often call $F$ and $M(F)$ with this property non-degenerate
as well.
\eDe

In many interesting cases
 $L$ has an element $Y$  with the following properties:

\romanparenlist
\begin{enumerate}
\item 
  $\ad Y$ is diagonizable,

\item 
  the spectrum of $\ad Y$ is real and discrete and eigenspaces
  are finite-dimensional,

\item 
  the number of negative eigenvalues is finite.
\end{enumerate}
\noindent
Such an element
  is called a \emph{hypercharge operator}.  It defines the
  \emph{triangular decomposition}:
  \begin{displaymath}
    L=L_- + \fg_0 +L_+ \, ,
  \end{displaymath}
where $L_-$ is the sum of eigenspaces of $\ad Y$ with negative
eigenvalues, $\fg_0$ is the
$0$-th
eigenspace and $L_+$ is
the product of eigenspaces with positive eigenvalues.

We let
$L_0 =\fg_0 +L_+$.  Both $L_+$ and $L_0$ are  open subalgebras of $L$.\\

Suppose that $Y|_F$ is a scalar operator (which is true
if $F$ is finite-dimensional irreducible $L_0$-module).
An eigenvector of $Y$ in $\Sing M(F)\,\backslash\, \Sing F$
is called a \textit{ non-trivial singular vector} of $M(F)$.
Denote by $V(F)$ the quotient of the $L$-module $M(F)$ by the
submodule generated by all non-trivial singular vectors,
that are eigenvectors of $Y$. Their eigenvalues are
necessarily different from those of trivial singular eigenvectors
so the map of $F$ to $V(F)$
is injective. We will often identify $F$ with its image in $M(F)$ or $V(F)$
depending on the module under consideration.

Clearly $V(F)$ could be irreducible even if $M(F)$
is not, which often happens when $M(F)$ is degenerate,
but not always. To study this we are to look at
singular vectors in $V(F)$.

Elements of $\Sing V(F)$ are called \emph{secondary singular
vectors}.  The image of $\Sing F \subset F $ in $V(F)$ lies
in $\Sing V(F)$ and
is called the subspace of \emph{trivial secondary singular
vectors}.

\bPr
  \label{prop:1.3}
Let $L$ be a linearly compact Lie superalgebra with a hypercharge
operator $Y$.  Then

\alphaparenlist
\begin{enumerate}
\item 
  Any finite-dimensional $L_0$-module $F$ is continuous.

\item 
  If $F$ is a finite-dimensional $L_0$-module, then $M(F)$ is in
  $\Pw(L,L_0)$.

\item 
  In any irreducible finite-dimensional $L_0$-module $F$ the
  subalgebra $L_+$ acts trivially.

\item 
  If $F$ is an irreducible finite-dimensional $L_0$-module, then
  $M(F)$ has a unique maximal submodule.

\item 
  Denote by $I(F)$ the quotient by the unique maximal
  submodule of $M(F)$.
  The map $F \mapsto I(F)$ defines a bijective correspondence
  between irreducible finite-dimensional $\fg_0$-modules and
  irreducible $L$-modules from $\Pw(L,L_0)$, the inverse map
  being $V \mapsto \Sing V$.

\item
  The $L$-module $M(F)$ is irreducible if and only if the $L_0$-module
  $F$ is
  irreducible and $\Sing M(F)  =F$.

\item 
  If the
    finite-dimensional
                       $L_0$-module $F$ is irreducible,
  and all its secondary singular vectors are trivial, then
  the $L$-module $V(F)$ is irreducible
  (and coincides with  $I(F)$).

\item 
  If $\tilde{S}$ is an irreducible $L_0$-submodule of $M(F)$ and
  $S$ is the $L$-submodule of $M(F)$ generated by $\tilde{S}$, then $S$ is
  irreducible iff $\Sing S = S \cap \Sing M(F)=\tilde{S}$.

\end{enumerate}
\ePr

\begin{proof}
  Let $F$ be a finite-dimensional $L_0$-module and let $v$ be a
  generalized eigenvector of $Y$ with eigenvalue $\lambda$.  If $
 a$ is an eigenvector of $\ad Y$ with eigenvalue $j$, it follows
 that $a(v)$ is a generalized eigenvector with eigenvalue
 $\lambda +j$.  Hence $v$ is annihilated by all but finitely many
 eigenspaces of $\ad Y$, proving~(a).

One has a filtration of $L$ by open subspaces given by $L_j=($
product of eigenspaces of $\ad Y$ with eigenvalues $\geq j)$.  Now
(a) and the proof of Proposition~\ref{prop:1.1}b  prove~(b).

Similarly,
 one shows that all elements from $L_+$ act on $F$ as nilpotent
 operators and therefore, by the superanalog of Engel's theorem,
they annihilate a non-zero vector.  Since the space spanned by these
 vectors is $L_0$-invariant, it coincides with $F$,
 which proves~(c).

If $F$ is an irreducible $L_0$-module, it is actually an
irreducible $\fg_0$-module (with $L_+$ acting trivially) on which therefore $Y$ acts as a scalar, let
it be $y_0$.  Then clearly  $Y$ acts diagonally on $M(F)$ in such a way that
$F$ coincides with its eigenspace for the eigenvalue $y_0$,
and $\Re(y)<\Re(y_0)$ for any other eigenvalue $y$ of $Y$ on $M(F)$.
This implies~(d). Then~(e),~(f)  and~(g) follow.

The statement~(h) follows from~(f), as soon as
we notice that the inclusion of $L_0$-modules
$\tilde{S} \subset M(F)$ induces the morphism  of $L$-modules
$M(\tilde{S}) \arr M(F)$ and the map is injective by PBW theorem,
therefore $S=M(\tilde{S})$.
\end{proof}

One has the following well-known corollary of
Proposition~1.3.

\begin{Corollary}
  \label{cor:1.4}
An $L$-module $M(F)$ is irreducible (hence non-degenerate) if and only if the
$\fg_0$-module $F$ is irreducible and $M(F)$ has no non-trivial
singular vectors
\end{Corollary}

\begin{remark}
\textup{The correspondence defined by Proposition~1.3e provides
 the classification of irreducible modules
of the category $\Pw(L,L_0)$.
For the non-degenerate of those modules the definition of
$M(F)$  supplies the
construction, and
Proposition~1.3g gives a construction of
the degenerate
modules having only trivial secondary singular vectors, provided
that one has a description of singular vectors.  }
\end{remark}

\section{Construction and basic properties of $E(3,6)$.}

One way to construct $E(3,6)$ is via its embedding into $E(5,10)$.
We describe first the geometric construction of $E(5,10)$ from
[CK1], Section~5.3 or [K1], Section~5.

Let $x_1, \ldots, x_5$  be even variables with  $ \deg x_i = 2$,
and let
$S_5$ be the Lie algebra of  divergence zero vector fields in these
 variables. Let $\dst\Om^1(5)$ be the space of closed differential
$2$-forms in these variables. Choosing degrees of the variables
and the degree of $\dst$ determines a
$\ZZ$-grading in vector fields and differential forms. We let the degree of
$\dst$ be $-5/2$, so that deg $\dvar i=-1/2$.

The Lie superalgebra $E(5,10)$ is constructed as follows:  $E(5,10)_{\bar{0}}\simeq S_5$ as a Lie algebra,
$E(5,10)_{\bar{1}}\simeq\dst\Om^1(5)$
as an $S_5$-module. The bracket on $E(5,10)_{\bar{1}}$ is
defined as the exterior product
of  differential forms which is a closed $4$-form identified with
the vector field whose contraction with the volume form produces
this $4$-form.  This construction gives a $\ZZ$-grading in $E(5,10)$ that we will
call the \emph{consistent} $\ZZ$-grading (since its even and odd
numbered pieces are comprised of even and odd elements, respectively).

In order to make explicit calculations
we will use the following notations:
\begin{displaymath}
\ddind jk  := \dvar{j} \wedge \dvar{k}, \quad
\dpind i   := \dprt /\dprt x_i  .
\end{displaymath}
We assume that the volume form is $\dvar{1}\wedge \cdots\wedge \dvar{5}$.
Now an element $A$ from $E(5,10)_{\bar{0}}=S_5$ can be written as
\[ A=\sum_i a_i \dpind{i}\, , \s \mx{ where } a_i \in \CC \left[\left[
  x_1,\ldots,x_5\right] \right] ,\s
 \sum_i \dpind{i}a_i =0 \, ,
\]
and an element $B$ from $E(5,10)_{\bar{1}}$ is of the form
\[ B=\sum_{j,k} b_{jk}\ddind jk \, , \hbox{ where }
b_{jk} \in \CC [[ x_1 , \ldots , x_5 ]], \, dB=0 \, .
\]
In particular the brackets in $E(5,10)_{\bar{1}}$
can be computed using bilinearity and the rule
\[ [a\ddind jk ,b\ddind lm ]= \ep_{ijklm} ab \dpind{i}
\]
where
$\ep_{ijklm}$ is the sign
of the permutation $(ijklm)$ when
$\{i,j,k,l,m\}=\{1,2,3,4,5\}$ and zero otherwise. \\

By definition ([K1], Example 5.4) the
algebra $E(3,6)$ is a consistently $\ZZ$-graded simple linearly compact
Lie superalgebra such that

\bEaz
\gind{0} &\simeq &  sl(3)+ sl(2) + gl(1) ,\\
\gind{-1} &\simeq & \CC^3\otimes \CC^2\otimes \CC(-1), \\
\gind{-2}&\simeq &
             \CC^3\otimes 1 \otimes \CC(-2), \\
\gind{-3}&\simeq & 0,\\
\gind{1} &\simeq & S^2\CC^3\otimes \CC^2\otimes \CC(1) +
             {\CC^3}^*\otimes \CC^2\otimes \CC(1).
\eEaz

We may construct $E(3,6)$ as a subalgebra of $E(5,10)$ as follows
(we slightly modify  the construction from [CK1]).  Consider
the \emph{secondary grading in $E(5,10)$} defined by the
conditions:
\begin{align}
\label{2nd-grad}
\bAr{ll}
\deg x_1 = \deg x_2 =  \deg x_3 = 0, &
\deg \dpind{1}
=\deg \dpind{2} =\deg \dpind{3} = 0,\\
\deg x_4 =  \deg x_5 =1, \s &
\deg \dpind{4} =\deg \dpind5 =-1\, ,\,   \deg \dst = -1/2 \, .
\eAr
\end{align}

\bPr
\cite{CK1} For the secondary grading of $E(5,10)$,
the zero-degree subalgebra
is the Lie superalgebra $E(3,6)$  .
The consistent $\ZZ$-grading in $E(3,6)$ is
induced by the consistent grading of $E(5,10)$.
\ePr

As a result we have for $L = E(3,6)$ the following description
of the first three pieces of its consistent $\ZZ$-grading $L =\Pi_{j \geq -2} \fg_j$:
\begin{displaymath}
\gind{-2}= \lsp{\dpind{i}, i=1,2,3}, \quad
\gind{-1} =  \lsp{\ddind ij ,i=1,2,3,j=4,5}.
\end{displaymath}
We shall use the following basis of $\fg_0 = s \ell (3) + s \ell
(s) + g\ell (1)$:
\begin{eqnarray*}
%
&&h_1=x_1\dpind 1-x_2\dpind 2, \quad
h_2=x_2\dpind 2-x_3\dpind 3, \quad
e_1=x_1\partial_2, \quad
e_{12}=x_2\partial_3, \quad e_3=x_1\partial_3,\\
&&f_1=x_2\partial_1, \quad
f_2=x_3\partial_2, \quad
f_{12}=x_3\partial_1, \quad
h_3=x_4\dpind 4-x_5\dpind 5, \quad
e_3=x_4\dpind 5,  \quad f_3 =x_5\dpind4,\\
&&Y=\frac{2}{3}(x_1\dpind 1+x_2\dpind 2+x_3\dpind 3)-(x_4\dpind 4+x_5\dpind 5).\\
\end{eqnarray*}
Here $s\ell (3)$ (resp. $s\ell (2)$) is spanned by elements
involving indeterminates $x_i$ with $i=1,2,3 $ (resp. $4,5$) and
$g\ell (1)=\CC Y$.
We use the element $Y$ as the hypercharge operator
and
we fix the
standard Cartan subalgebra $\Hw=\lsp{h_1,h_2,h_3,Y}$ and
the standard Borel
subalgebra $\Bw=\Hw \oplus  \Nw $, where $\Nw =\langle e_i (i=1,2,3), e_{12} \rangle$, of $\gind
0$.  Note that the eigenspace decomposition of $3Y$ coincides
with the consistent $\ZZ$-grading of $E(3,6)$.  (Incidentally,
$E(5,10)$ has no hypercharge operators.)

The algebra $ E(3,6)$ is generated by
$\gind{-1}, \gind 0, \gind 1$ \cite{CK1}; moreover it is generated
by $\gind 0$ and the following three elements $e_0, e'_0, f_0$:
\begin{eqnarray*}
f_0 &=&\ddind 14,\\
e'_0 &=& x_3\ddind 35,  \\
e_0 &=& x_3\ddind 25-x_2\ddind 35+2 x_5\ddind 23 \, ,
\end{eqnarray*}
where the element $f_0$ is the highest weight vector of the
$\fg_0$-module $\gind{-1}$, while $e'_0, e_0$ are the lowest
 weight vectors of the $\fg_0$-module $\gind{1}$, and one has:
 \begin{eqnarray}
   \label{eq:2.2}
   [e'_0,f_0] &=& f_2, \\
   \label{eq:2.3}
   [e_0,f_0] &=& \frac{2}{3}h_1+\frac{1}{3}h_2- h_3 - Y =: h_0.
 \end{eqnarray}

So the elements $\{h_i,e_i, f_i (i=0,1,2,3),e'_0\} $ generate $E(3,6)$.  We call them the generalized
Chevalley generators of $E(3,6)$ (since apart from (\ref{eq:2.2})
they satisfy the relations satisfied by the ordinary Chevalley
generator of a semisimple Lie algebra).

The above observations give the following proposition.

\begin{Proposition}
  \label{prop:2.2}
The elements $e_i (i=0,1,2,3)$ and $e'_0$ generate $ \Nw + L_+$.
Consequently, a $\fg_0$-highest weight vector $v$ of a
$E(3,6)$-module is singular iff
\begin{displaymath}
  e_0 \cdot v =0, \,\, e'_0 \cdot v =0 \, .
\end{displaymath}

\end{Proposition}

In order to see the action of $\fg_0$ on the space
$\gind{-1} =  \lsp{\ddind ij ,i=1,2,3,j=4,5}$
more clearly we write
\begin{displaymath}
  \dst^+_i := \dst_{i4} \qquad \dst^-_i := \dst_{i5} \, .
\end{displaymath}
and we define
$\lag^{\pm}_{-1}=\lsp{\dst^{\pm}_1,\dst^{\pm}_2,\dst^{\pm}_3}.$
We also use the following shorthand notations for
the elements
from $\La\lsp{\dst^-_1,\dst^-_2,\dst^-_3}$:
\begin{displaymath}
\dst^-_{ij} := \dst^-_i \cdot \dst^-_j, \qquad
\dst^-_{ijk} := \dst^-_i \cdot \dst^-_j\cdot \dst^-_k \, ,
\end{displaymath}
and similarly for the ``$+$''-type. We let
$\La^{\pm}:=\La \langle\dst_1^{\pm},\dst_2^{\pm},\dst_3^{\pm}\rangle$.

Consider the following abelian subalgebras of $\fg_1$, normalized
by $s\ell (3)$:
\begin{displaymath}
  \fg^+_1 = \langle x_id_{j5} + x_j d_{i5} | i,j=1,2,3 \rangle \,
  ,\,\, \fg^-_1 =\langle x_i d_{j4}+x_jd_{i4} | i,j,=1,2,3 \rangle \,
  ,
\end{displaymath}
and let
\begin{displaymath}
  S(3)^{\pm} = \fg^{\pm}_{-1} + s\ell (3) + \fg^{\pm}_1 \, .
\end{displaymath}
It is easy to see that $S(3)^{\pm}$ are subalgebras of $\fg$
isomorphic to the simple Lie superalgebra $S(3)$ of
divergenceless vector fields in three anticommuting
indeterminates.  Note that
\begin{equation}
  \label{eq:2.4}
  [\fg^{\pm}_{-1}, \fg^{\mp}_1] =0 \, .
\end{equation}


One can check that
$E(3,6)_{\bar{0}}\simeq W_3 + \Om^0(3)\otimes sl(2)$ and
$E(3,6)_{\bar{1}}\simeq \Om^1(3)\otimes\CC^2$.
Here the first
isomorphism maps
 $D \in W_3$  to $ D-\frac{1}{2} \DIV(D)(x_4\dpind 4+x_5\dpind 5)$ and
is identical on the second summand.
The second isomorphism could be chosen according to the following
formula (which differs from the one in [CK1])
\[
f\dvar i\cdot \ep_a \arr -\dst(f\dvar i \cdot x_{a+3}),
i=1,2,3,\s a=1,2 \, ,
\]
where $\ep_a, a=1,2$, is the standard basis in $\CC^2$.
Of course it is possible to define
brackets in $E(3,6)$ in terms of these isomorphisms
and this gives the construction of $E(3,6)$ from \cite{CK1}.


\section{Lemmae about $sl(3)$-modules}
\label{sec:3}

>From now on we let $L = E(3,6)$.  As before, we use notation
$L_-=\oplus_{j<0} \fg_j, L_+=\Pi_{j>0} \fg_j,  L_0 =\fg_0 + L_+$.
We shall use the
realization of this Lie superalgebra as a subalgebra of $E(5,10)$
described in Section~2.  As explained in Section~1, our first
main objective is to study irreducibility of the induced
$\fg$-modules
\begin{equation}
  \label{eq:3.1}
  M(V) = U(L ) \otimes_{U(L_0)}
V \cong U (L_-) \otimes V \, ,
\end{equation}
where $V$ is a finite-dimensional irreducible $\fg_0$-module
extended to $L_0$ by letting $\fg_j$ for $j>0$ acting
trivially.  The isomorphism in (\ref{eq:3.1}) is an isomorphism
of $\fg_0$-modules, which can be used to define the action of
$L$ on $U(L_-) \otimes V$ (in particular $L_-$acts by left
multiplication).

Recall that $\fg_0 = s \ell (3) \oplus s \ell (2) \oplus g\ell
(1)$, where
\begin{displaymath}
  s\ell (3) = \langle x_i\partial_j | 1 \leq i,j \leq 3 \rangle
  \cap \fg_0, \quad
  s\ell (2) = \langle x_i\partial_j | i,j =4,5 \rangle  \cap \fg_0\, .
\end{displaymath}
Hence it is important to have a \emph{model} for $s\ell (3)$,
i.e.,~an $s\ell (3)$-module in which every finite-dimensional
irreducible $s\ell (3)$-module appears exactly once.  Note that
$s\ell (3)$ acts on the polynomial algebra $\CC [\partial_1,
\partial_2, \partial_3, x_1,x_2,x_3]$ in a natural way (by
derivations $x_i\partial_j (x_k)= [x_i\partial_j,x_k] =
\delta_{jk} x_i,\,\, x_i\partial_j (\partial_k) =
[x_i\partial_j,\partial_k] = -\delta_{ik} \partial_j$), so that
the element $P:=\partial_1 x_1 + \partial_2 x_2 +\partial_3x_3$
is annihilated.  Denote by $\Mw$ the quotient of this polynomial
algebra by the ideal generated by $P$, with the induced action of
$s\ell (3)$.

\bLe The $s\ell (3)$-module $\Mw$ is a model.  The irreducible
$s\ell (3)$-module with highest weight $(m,n)$ appears in $\Mw$ as
the bigraded component
%
%
\[\lsp{ \dprt_1^{a_1}\dprt_2^{a_2}\dprt_3^{a_3}x_1^{b_1}x_2^{b_2}x_3^{b_3}|
\,\,\sum a_i=m,\,\sum b_i=n}.
\]
The highest weight vector of this submodule is
$\dprt_3^{n}x_1^{m}$.
\eLe

\begin{proof}
  Let $U$ denote the subgroup of the group $G=SL(3,\CC)$
  consisting of upper triangular matrices with $1$'s on the
  diagonal.  It is well-known that in the space $\CC [G/U]$ of
  regular functions on $G/U$ all irreducible finite-dimensional $
 G$-modules occur exactly once.  On the other hand, $G/U$ is
 isomorphic to the orbit of the sum of highest weight vectors in
 the $G$-module $\CC^3 \oplus \CC^{3*}$ and this orbit is the
 complement to $0$ in the quadric $\sum_i x_i\partial_i=0$, where $x_i$
 (resp.~$\partial_i$) are standard coordinates on $\CC^3$
 (resp.~$\CC^{3*}$).  Since this quadric is a normal variety, we
 conclude that the $G$-module $\CC [G/U ]$ is isomorphic to $\Mw$.
 The lemma follows.

\end{proof}



Thus, every $L$-module $M(V)$ is contained in $U(L_-)\otimes
\Mw \otimes T$, where $T$ is a (finite-dimensional irreducible) $
s\ell (2)$-module.  We shall use the following shorthand
notation:
\begin{displaymath}
  u[m]t=u \otimes m \otimes t \in U(L_-) \otimes \Mw \otimes T \, .
\end{displaymath}
(This notation also reminds one that elements of $\Mw$ are
cosets.)  We shall mark the elements $\partial_i \in \fg_{-2}$ by
a hat in order to distinguish them from the elements $\partial_i$
used in the construction of $\Mw$.  We let $\Sym=\CC
[\hat{\partial}_1, \hat{\partial}_2, \hat{\partial}_3]$.

We shall consider the tensor product $U(L_-) \otimes \Mw$ of
associative algebras.  It is a $U(s\ell (3))$-module with the
usual action on the tensor product.  Hence we may consider the
smash product
\begin{displaymath}
  U=(U(L_-)\otimes \Mw) \# U (s\ell (3)) \, .
\end{displaymath}
This is an associative algebra which acts on $U(L_-) \otimes
\Mw$ in the obvious way (elements from $U(L_-) \Mw$ act by left
multiplication).

The algebra $U(L_-) \otimes \Mw$ contains a commutative $U
(s\ell (3))$-invariant subalgebra $\Sym \otimes \Mw$.  In the
following proposition and further we shall denote by $\wt_3 v$ the
$s\ell (3)$-weight of a vector~$v$.

\bPr\label{P:Dstr}
Consider the following elements of $\Sym \otimes \Mw$:
$$\bAr{rll}
\bar{D}_1&=&\dhind1 [x_1]+\dhind2 [x_2]+\dhind3 [x_3],\\
\bar{D}_2&=&\dhind2[\dpind3] -\dhind3[\dpind2],\\
\bar{D}_3&=&\dhind3 [1] = \hat{\partial}_3.
\eAr$$
Any $sl(3)$-highest weight vector $\bar{w}$
in
$
\Sym\otimes \Mw
$
can be uniquely written as
$$
\bar{w}=\sum_{\al,m,n} \bar{c}_{\al;m,n} \bar{D}_3^{\al_3}\bar{D}_2^{\al_2}\bar{D}_1^{\al_1}
 [\dprt_3^{n}x_1^{m}].
$$
If $\wt_3 \bar{w}=(a,b)$ then
$(m,n)=(a,b)-(\al_2,\,\al_3)$ for non-zero $\bar{c}_{\al;m,n}$.
\ePr

\begin{proof}
  It follows from [Sh] that the algebra of $U$-invariants for the
  action of $SL(3)$ on the algebra $A:=\CC [\hat{\partial}_1,
  \hat{\partial}_2, \hat{\partial}_3, \partial_1,  \partial_2,
  \partial_3, x_1, x_2, x_3]$ is generated by six algebraically
  independent elements:
  $\bar{D}_0:= \sum_i \partial_i x_i , \bar{D}_1, \bar{D}_2,
  \bar{D}_3, \partial_3$ and $x_1$.  Since
  $\Sym \otimes \Mw \cong A/(\bar{D}_0)$
  by Lemma~3.1, the proposition follows.
\end{proof}

Suppose $\bar{w} \in \Sym\otimes M$,
where $M\subset \Mw$ is an irreducible $sl(3)$-module
generated by the highest vector
$[\dprt_3^{\nu_2}x_1^{\nu_1}]\in\Mw$.
Then in the above formula we should have
$(\nu_1,\nu_2)=(m,n)+(\al_1,\al_2)$.
So the weights of $s\ell (3)$-highest weight vectors
in $\Sym\otimes M$ are
\[
(a,b)=(\nu_1,\nu_2)-(\al_1,\,\al_2)+(\al_2,\,\al_3)
\]
where $\al_i\leq \nu_i$, $i=1,2.$



Suppose now that an $sl(3)$-module $M$ is given
as an abstract finite-dimensional module. We would like to extend
our description of the highest weight vectors in $\Sym \otimes \Mw $
to  $\Sym \otimes M$. Note that $\Sym \otimes M$ is an $\Sym$-module
via the left multiplication, and also a $U(s\ell (3))$-module
with the usual action on tensor product, so that we have the
action of $\Sym \# U(s \ell (3))$ on $\Sym \otimes M$.

Let $h=h_1+h_2+1$.  Instead of $\bar{D}_i$, we define the  operators $D_i$
from $\Sym \# U( sl(3))\subset U$ as follows (as before, we drop the tensor signs):
  \begin{align}                                       
D_1&=\dhind1\,h_1h +
\dhind2\, f_{12}h +
\dhind3\,(f_3h_1+f_2f_1),\notag
\\
D_2&=\dhind2\,h_2+\dhind3\, f_2,     \label{E:DD}\\
D_3&=\dhind3 \notag.
\end{align}
The action of the operators $D_i$
on $\Sym \otimes \Mw $ is related to the left multiplication by the
$\bar{D}_i$ by the formulae\\
  \begin{alignat}{2}                                
D_1(s[\dpind3^q x_1^p]) &= &
\;p(p+q+1)\; &\bar{D}_1(s[\dpind3^q x_1^{p-1}]),\notag
\\
D_2(s[\dpind3^q x_1^p]) &= &\quad
q\; &\bar{D}_2 (s[\dpind3^{q-1} x_1^p]),
                                                       \label{E:D-DD}\\
D_3(s[\dpind3^q x_1^p]) &= & &\bar{D}_3 (s[\dpind3^q x_1^p]).
\notag
\end{alignat}
Note that equations~(3.2) represent the defining property of the
operators $D_i$, and (3.1) is a solution of these equations.

\bPr
\label{P:DD-prop} 

  The operators $D_i$ commute with each other,
  and  while acting on $\Sym \otimes \Mw $ the
  operator $D_i$ commutes with $\bar{D}_j$ for $j< i$.
\ePr

\begin{proof}
This is not difficult to check by a straightforward calculation
(see below).
\end{proof}

In the following $A^{[n]}:=A(A-1)\cdots(A-n+1)$.

\bPr                                      \label{P:Dstr-tilde}
Let $M$ be an irreducible $s \ell (3)$-module
with the highest weight vector $m_0$.
Any highest weight vector in  $\Sym\otimes M$
can be written uniquely as a linear combination of the form
\[
w=\sum_{\al} c_{\al}
 D_1^{\al}D_2^{\al_2}D_3^{\al_3}
  m_0 .
\]
\ePr
\begin{proof}

Pick  a monomorphism $\mu:M\arr \Mw$
such that
$\,\,\,\mu(m_0)=[\dprt_3^{\nu_2}x_1^{\nu_1}]\in\Mw$ for a highest
weight vector $m_0$ of $M$.
Clearly, by Proposition~3.3 and equation~(3.3), for the
expression for $\bar{w}$ given by  Proposition~\ref{P:Dstr}
we have:
\begin{equation}                       \label{E:hi-v-tilde}
w=\sum_{\al,m,n} c_{\al;m,n}
D_1^{\al_1}D_2^{\al_2}D_3^{\al_3}
 m_0 , \quad \text{ where }\qquad \mu\,( w)=\bar{w}
\end{equation}
and
\begin{equation}                       \label{E:hi-v-coeff} 
(\nu_1)^{[\al_1]}(\nu_1+\nu_2+1)^{[\al_1]}(\nu_2)^{[\al_2]}\,
c_{\al;m,n}=\bar{c}_{\al;m,n}
\end{equation}

\end{proof}

Any element $v$ of $ S \otimes \Mw$ can be written uniquely in
the form
\begin{displaymath}
  v=\sum_{\alpha \in \ZZ^3_+} \hat{\partial}^{\alpha_1}_1
  \hat{\partial}^{\alpha_2}_2 \hat{\partial}^{\alpha_3}_3
  t_{\alpha} \equiv \sum_{\alpha} \hat{\partial}^{\alpha}
  t_{\alpha} \, , \hbox{  where  } t_{\alpha} \in \Mw \, .
\end{displaymath}
We define $\ell ht\,  v = \hat{\partial}^{\sigma} t_{\sigma}$, where
$\sigma$ is the lexicographically highest element of the set\break $\{
\alpha \in \ZZ^3_+ | t_{\alpha} \neq 0 \}$.  It is immediate to
see
\begin{equation}
  \label{eq:3.6}
  \ell ht\,  \bar{D}^{\alpha} [\partial^n_3 x^m_1]
     =\hat{\partial}^{\alpha}
     [\partial^{n+\alpha_2}_3 x^{m+\alpha_1}_1] \, .
\end{equation}

\begin{Proposition}
  \label{prop:3.5}
  $\ell ht \, D^{\alpha} =\hat{\partial}^{\alpha} h^{[\alpha_1]}
  h_1^{[\alpha_1]} h^{[\alpha_2]}_2$.

\end{Proposition}

Using (\ref{eq:3.6}), we get the following corollary.

\begin{Corollary}
  \label{cor:3.6}
  \begin{displaymath}
\ell ht \, D^{\alpha} [\partial^q_3  x^p_1]
 = p^{[\alpha_1]}
(p+q+1)^{[\alpha_1]}
 q^{[\alpha_2]} \ell ht \, \bar{D}^{\alpha}
   [\partial^{q-\alpha_2}_3  x^{p-\alpha_1}_1] \, .
  \end{displaymath}

\end{Corollary}

The proof of Proposition~{\ref{prop:3.5}} is based on several
lemmas, which also establish some properties of the operators
$D_i$ used in the sequel.

\begin{Lemma}
  \label{lem:3.7}
$D_i$ commute with each other.

\end{Lemma}

\begin{proof}

Let $A=\hat{\partial}_1 h_1 + \hat{\partial}_2 f_1$, $B = f_{12}
h_1 + f_2 f_1$ so that $D_1=Ah+\hat{\partial}_3 B$.  We see that
$ [A,\hat{\partial}_2]=0$, $[A,h_2]=0, [A,\hat{\partial}_3]=0$
and $[A,f_2]=\hat{\partial}_1 f_2 -\hat{\partial}_2 f_{12}$.
Therefore
\begin{displaymath}
  [A,D_2] = \hat{\partial}_1 \hat{\partial}_3 f_2 -
  \hat{\partial}_2 \hat{\partial}_3 f_{12} \, .
\end{displaymath}
Also $[h,D_2]=0$, $[\hat{\partial}_3 , D_2]=0$ and
\begin{eqnarray*}
  [B, \hat{\partial}_2] &=& \hat{\partial}_2 f_{12}
     - \hat{\partial}_1 f_2 \, , \\ {}
  [B, h_2] &=& B \, , \\ {}
  [B, \hat{\partial}_3] &=& -\hat{\partial}_1 h_1
     -\hat{\partial}_2 f_1 = -A \, , \\ {}
  [B,f_2] &=& [f_{12}, f_2] =0 \, ,
\end{eqnarray*}
thus
\begin{displaymath}
  [B,D_3] = (\hat{\partial}_2 f_{12} -\hat{\partial}_1 f_2)
  h_2 + \hat{\partial}_2 B -A f_2 =(\hat{\partial}_2 f_{12} -
  \hat{\partial}_1 f_2 )h \, .
\end{displaymath}
We conclude that $[D_1,D_2] =0$.  We have computed that
$[A,\partial_3] =0$, $[B,\partial_3] =-A$, and $[h,\partial_3]
=\partial_3$, so $[D_1 , D_3]=0$.  Clearly $[D_2,D_3]=0$ as well.

\end{proof}

The following lemma could be applied either to $D_1$ or $D_2$.
It provides quite a nice expression for $D^k_i$ (here we will
need only the lexicographically highest term of the sum but later
we will also use the second one).

\begin{Lemma}
  \label{lem:3.8}
Let $D=ah+\partial b$ where
\begin{eqnarray*}
  \begin{array}{ll}
[a,b]=0 , & [\partial ,b] =+a , \,\, [\partial ,a] =0 \, , \\ {}
[h,b]=-2b , & [h,a]=-a , \,\, [h, \partial] =\partial  \, .
  \end{array}
\end{eqnarray*}
Then
\begin{displaymath}
  D^k = \sum^k_{m=0} \binom{k}{m} \partial^m b^m a^{k-m}
  (h-m)^{[k-m]} \, .
\end{displaymath}

\end{Lemma}

\begin{proof}

The formula is easily proven by induction on $k$.

\end{proof}

One can easily check that $D_1$ with $a=A, b=B$ satisfies the
above lemma, as well as $D_2$ with $a=\hat{\partial}_2, b=f_2$
and $A$ with $a=\hat{\partial}_1, b=f_1$.  This makes it easy to
compute the following lexicographically highest terms.

\begin{Corollary}
  \label{cor:3.9}

$\ell ht \, D^k_1 = \hat{\partial}^k_1 h^[k] h_1^{[k]}, \ell ht \, D^{\ell}_2 =
   \hat{\partial}_2^{\ell} h_2^{[\ell]}$.
\end{Corollary}

\begin{Lemma}
  \label{lem:3.10}
Let $D_2 \{ +m \} = \hat{\partial}_2 (h_2 +m) + \hat{\partial}_3
f_2$.  Then $[D_2 , \hat{\partial}_3] = [D_2
,\hat{\partial}_1]=0$ and $D_2 \hat{\partial}_2=\hat{\partial}_2
D_2 \{ +1 \}$.
\end{Lemma}

\begin{proof}
This is a straightforward calculation.
\end{proof}

\begin{Corollary}
  \label{cor:3.11}
If $\ell ht \, f = \hat{\partial}^{\alpha} u$ then $\ell ht (D_2f) =
\ell ht (\hat{\partial}^{\alpha} D_2 \{ +\alpha_2 \} u)$.
\end{Corollary}

\begin{proof}[Proof of Proposition~\ref{prop:3.5}.]

We can apply Corollary~\ref{cor:3.11} to the situation of the
proposition:
\begin{displaymath}
  \ell ht (D^{\ell}_2 D^k_1)
      = \ell ht (\hat{\partial}^k_1 (D_2)^{\ell} h^{[k]}
      h^{[k]}_1)
      = \hat{\partial}^k_1 \hat{\partial}^{\ell}_2
      h^{[\ell]}_2 h^{[k]} h^{[k]}_1 \, .
\end{displaymath}
The formula of the proposition follows.

\end{proof}

Now we describe the $sl(3)$-highest weight vectors in
$
\La^{\pm}\otimes F
$.
We here omit $\pm$ because the results are exactly the same for
``$+$'' and for ``$-$''.

\bLe\label{la-high} Let
$F\subset \Mw$ be an irreducible finite-dimensional $sl(3)$-module
with  highest weight $(p,q)$.
For the $sl(3)$-highest weight element
$u \in \La\lsp{\ddind1{},\ddind2{},\ddind3{}}\otimes F$
of weight  $(m,n)=(p,q)+\de$,  there are
the following possibilities (up to a constant factor):\\
\begin{tabular}{llcl}
{\rm(00)$'$:} & $\de=(\,0,\,0)$ & and &
$
u= [\dpind 3^n x_1^m] \, ,
$\\
{\rm(+0):} & $\de=(+1,\,0)$  & and &
$
u= \ddind 1{}[\dpind 3^{n} x_1^{m-1}]\,,
$\\
{\rm(--+):} & $\de=(-1,\,1)$  & and &
$
u= (\ddind 1{}[x_2]-\ddind 2{}[x_1]) \,
[\dpind 3^{n-1} x_1^m]\,,
$\\
{\rm(0--):} & $\de=(\,0,-1)$ & and &
$
u=(\ddind1{}[\dpind1]+\ddind2{}[\dpind2]+\ddind3{}[\dpind3]) \,
[\dpind 3^n x_1^m]\,,
$
\\
{\rm(0+):}  &  $\de=(\,0,+1)$ & and &
$
u= \ddind1{2}[\dpind 3^{n-1} x_1^{m}]\,,
$
\\
{\rm(--0):} & $\de=(-1,\,0)$ & and &
$
u= (\ddind1{2}[x_3]+\ddind 3{1}[x_2]+\ddind 2{3}[x_1]) \,
[\dpind 3^n x_1^m]\,,
$
\\
{\rm(+--):} &  $\de=(\,1,-1)$ & and &
$
u= \ddind1{}
(\ddind2{}[\dpind 2]+\ddind 3{}[\dpind 3])\,[\dpind 3^n x_1^{m-1}]\,,
$
\\
{\rm (00)$''$:} &  $\de=(\,0,\,0)$ &  and &
$
 u= \ddind1{23}[\dpind 3^n x_1^m] \, .
$
\end{tabular}\\
\eLe

\begin{proof}
This is standard and we leave the proof to the reader.
\end{proof}


In the following let
$\De^\pm:=\dst_1^{\pm}[\dpind1]+\dst_2^{\pm}[\dpind2]+\dst_3^{\pm}[\dpind3]$.

\bLe\label{f0-key} Let $F$ be an irreducible $sl(3)$-module
with highest weight $(p,q)$ and such than the action
of $L_+$ on $F$ is trivial.  If $u \in \La^+\otimes F$
is an $sl(3)$-highest weight vector of weight $(m,n)$
and $e'_0\cdot u=0$,
then
there are
the following possibilities for $u$ (up to a constant factor):\\
{\rm (T0):} $(m,n)=(p,q)$ and
$
u=\,\, [\dpind3^{q}x_1^{p}] \in F,
$
\\
{\rm(T1):}  $p\geq 0,\,q=0$, $(m,n)=(p+1,0)$ and
$
u=\,\,\dst_1^+\,[x_1^{p}]\,,
$
\\
{\rm (T2):} $p=0,\,q\geq 1$, $(m,n)=(0,q-1)$  and \\
\hspace*{.38in}$
u=(\dst_1^+[\dpind1] + \dst_2^+[\dpind2] + \dst_3^+[\dpind 3])
[\dpind3^{n}]=\De^+[\dpind3^{q-1}],
$
\\
{\rm (T3):} $(p,q)=(0,1)$, $(m,n)=(1,0)$ and
$
u= \dst_{12}^+[\dpind2]+\dst_{13}^+[\dpind 3]=\,\dst_1^+\De^+,
$
\\
{\rm (T4):} $(p,q)=(m,n)=(0,0)$ and
$
u= \, \,\dst_{123}^+
[1]\,.
$

\noindent  In particular in all cases except for (T0), either $p=0$ or $q=0$.
\eLe

\prf \/
We will write $u$ in the form provided by Lemma~3.12
and calculate $e'_0\cdot u$. We are to remember that $e'_0 (F)=0$
and the following relations
for the action of $e'_0$ on $\langle \dudind+1,\dudind+2,\dudind+3\rangle$
is important to have in mind.
\bEa\label{f'0-la+}
e'_0\cdot \dst_1^+ = f_2,\,\,
e'_0\cdot \dst_2^+  =- f_{12},\,\,
e'_0\cdot \dst_3^+  =0.
\eEa

Case (00): $(p,q)=(m,n)$ and
$ u= (c_0  + c_1 \dudind+{123})\, \dpind 3^n x_1^m,$
hence
\bEa
0&=&e'_0u=
 0+c_1\left( f_2\dudind+2\dudind+3\,[\dpind 3^n x_1^m]
-\dst_1^+(-f_{12})\dst_3^+\,[\dpind 3^n x_1^m]\right)=\nonumber\\
&=&c_1\left(0-n\,\dst_2^+\dst_3^+\,[\dpind2\dpind 3^{n-1} x_1^m]
- n\,\dst_1^+\dst_3^+\,[\dpind1\dpind 3^{n-1} x_1^m]
+m\,\dst_1^+\dst_3^+\,[\dpind 3^{n} x_1^{m-1}x_3]%
\right).\nonumber
\eEa We see that either $c_1=0$,
which gives us (T0), or $m=n=0$, which gives (T4).\\

Case (+0): $(p,q)=(m-1,n)$ and
$ u= \dst_1^+\, [\dpind 3^{n} x_1^{m-1}], $ so
\bEaz
0&=&e'_0u=
f_2 \,[\dpind 3^{n} x_1^{m-1}]=
 -n \,
[\dpind2\dpind 3^{n-1} x_1^{m-1}].
\eEaz
The solution exists for $n=0$, $m\geq 1$. This is (T1).\\

Case ($-$+): $(p,q)=(m+1,n-1)$ and
$ u=
(\dst_1^+[x_2]-\dst_2^+[x_1])
\,[\dpind 3^{n-1} x_1^m]\, .
$
We have:
\bEaz
0&=& e'_0u= f_2\, [\dpind 3^{n-1} x_1^{m}x_2]-
     (-f_{12})\, [\dpind 3^{n-1} x_1^{m+1}]=\\
&=&
-(n-1)\,[\dpind2\dpind 3^{n-2} x_1^{m}x_2]
                + [\dpind3^{n-1} x_1^{m}x_3]-\\
& &\quad-(n-1)\,[\dpind1\dpind 3^{n-2} x_1^{m+1}]
              +(m+1) \,[\dpind 3^{n-1} x_1^{m}x_3]\,.
\eEaz
This implies $-(n-1)=m+2$, but both
 $m$,$n$ are non-negative hence no solution is possible.\\

Case (0--): $(p,q)=(m,n+1)$ and
$ u= \Delta^+
\,[\dpind 3^n x_1^m]$. We  have:
\bEaz
0&=&e'_0u=
f_2\, [\dpind1\dpind 3^{n} x_1^{m}]
   -f_{12}\, [\dpind2\dpind 3^{n} x_1^{m}]=\\
&=&   -n\,[\dpind1 \dpind2\dpind 3^{n-1} x_1^{m}]
    +\,n\,[\dpind1 \dpind2\dpind 3^{n-1} x_1^{m}]
         -\,m\,[\dpind2\dpind 3^{n} x_1^{m-1}x_3]\\
    &=&-\,m\,[\dpind2\dpind 3^{n} x_1^{m-1}x_3]\,.
\eEaz
We conclude that $m=0$ and this gives (T2).\\

Case (0+): $(p,q)=(m,n-1)$ and
$ u=
\dudind+{12}\,
[\dpind 3^{n-1} x_1^{m}]\,,
$ therefore
\bEaz
0&=&e'_0u=f_2\dst_2^+ \,[\dpind 3^{n-1} x_1^{m}]
         -\dst_1^+(-f_{12})\,[\dpind 3^{n-1} x_1^{m}]\\
&=&          \dst_3^+ \,[\dpind 3^{n-1} x_1^{m}]
-(n-1)\dst_2^+ \,[\dpind2\dpind 3^{n-2} x_1^{m}]-\\
& &\quad-(n-1)\dst_1^+\,[\dpind1\dpind 3^{n-2} x_1^{m}]
            +m\,\dst_1^+\,[\dpind 3^{n-1} x_1^{m-1}x_3]\,.
\eEaz
The $\dst_3^+$-term shows that there are no solutions here.\\

Case (--0): $(p,q)=(m+1,n)$ and
$ u= (\dst_{12}^+[x_3]-\dst_{13}^+[x_2]+\dst_{23}^+[x_1])\,
[\dpind 3^n x_1^m].$
Then
\bEaz
0&=&e'_0u=
f_2\dst_2^+ \,[\dpind 3^{n} x_1^{m}x_3]
-\dst_1^+(-f_{12})\,[\dpind 3^{n} x_1^{m}x_3]-\\
& &\qquad\qquad-f_2\dst_3^+ \,[\dpind 3^{n} x_1^{m}x_2]
+(-f_{12})\dst_3^+ \,[\dpind 3^{n} x_1^{m+1}]\\
&=&           \dst_3^+ \,[\dpind 3^{n} x_1^{m}x_3]
       -n\dst_2^+ \,[\dpind2\dpind 3^{n-1} x_1^{m}x_3]
     -\,n\,\dst_1^+\,[\dpind1\dpind 3^{n-1} x_1^{m}x_3]+\\
& &\quad +m\, \dst_1^+ \,[\dpind 3^{n} x_1^{m-1}x_3^2]
   + \,n\,\dst_3^+ \,[\dpind2\dpind 3^{n-1} x_1^{m}x_2]
            - \,\dst_3^+ \,[\dpind 3^{n} x_1^{m-1}x_3]+\\
& &\qquad\quad +\,n\,\dst_3^+ \,[\dpind1\dpind 3^{n-1} x_1^{m+1}]
                     -(m+1)\,\dst_3^+ \,[\dpind 3^{n} x_1^{m}x_3]\,.
\eEaz
Now there is only one
$\dst_2^+$-term and this gives $n=0$.
Then we are left with only one
$\dst_1^+$-term and it gives $m=0$. Then we are left with a non-zero
$\dst_3^+$-term, so there are no solutions in this case.\\

Case (+$-$): $(p,q)=(m-1,n+1)$ and
$ u=
(\dst_{12}^+[\dpind 2]+\dst_{13}^+[\dpind 3])
\,[\dpind 3^n x_1^{m-1}].
$
Then
\bEaz
0&=&e'_0u=
 f_2\dst_2^+ \,[\dpind2\dpind 3^{n} x_1^{m-1}]
-\dst_1^+(-f_{12})\,[\dpind2\dpind 3^{n} x_1^{m-1}]
+f_2\dst_3^+ \,[\dpind 3^{n+1} x_1^{m-1}]=\\
&=&
\dst_3^+ \,[\dpind2\dpind 3^{n} x_1^{m-1}]
-n \,\dst_2^+ \,[\dpind2^2\dpind 3^{n-1} x_1^{m-1}]-
\,n\,\dst_1^+\,[\dpind1\dpind2\dpind 3^{n-1} x_1^{m-1}] +\\
& & \qquad+(m-1)\,\dst_1^+\,[\dpind2\dpind 3^{n} x_1^{m-2}x_3]
-(n+1)\,\dst_3^+\,[\dpind2\dpind 3^{n} x_1^{m-1}]\,.
\eEaz
Again there is only one
$\dst_2^+$-term and it gives $n=0$.
Then $\dst_1^+$-term gives $m=1$, and this gives (T3).\epf

\bLe\label{f1-key} Let $e'_1=x_3 d_{34} \in \fg_1$.
Let $F$ be an irreducible $sl(3)$-module
with highest weight $(p,q)$ and such than the action
of $L_+$ on $F$ is trivial.
If $u \in \La^- \otimes F$
is an $sl(3)$-highest weight vector of weight $(m,n)$
and $e'_1\cdot u=0$,
then
there are
the following possibilities for $u$ (up to a constant factor):\\
{\rm (T0):} $(m,n)=(p,q)$ and
$u=\, [\dpind3^{q}x_1^{p}] \in F,
$\\
{\rm(T1):}  $p\geq 0,\,q=0$, $(m,n)=(p+1,0)$ and
$u=\, \dst_1^-\,[x_1^{p}]\,,
$\\
{\rm (T2):} $p=0,\,q\geq 1$, $(m,n)=(0,q-1)$  and
$u=\,\De^-[\dpind3^{n}]\,,$
\\
{\rm (T3):} $(p,q)=(0,1)$, $(m,n)=(1,0)$ and
$
u=\, (\dst_{12}^-[\dpind2]+\dst_{13}^-[\dpind 3])
=\dst_1^-\De^-
\,,
$\\
{\rm (T4):} $(p,q)=(m,n)=(0,0)$ and
$u=\, \dst_{123}^-\,[1]\,.$
\eLe

\prf\/
As relations for $e'_1$
\bEa\label{f'1-la-}
e'_1\cdot \dst_1^-  =-f_2,\quad
e'_1\cdot \dst_2^-  =+f_3,\quad
e'_1\cdot \dst_3^-  =0
\eEa
differ only in the sign from the corresponding
relations~\eqref{f'0-la+} for $e'_0$, the calculations
above provide the proof. \epf

\begin{remark}
  \label{rem:3.14}
In Lemmas~3.6 and 3.7 we actually describe singular vectors of
induced $S(3)^{\pm}$-modules.
\end{remark}


\section{The highest $sl(3)$-weights of degenerate modules.}

We keep $L=E(3,6)$. Let
$V$ be a finite-dimensional irreducible $\gind 0$-module.
We are concerned with
singular vectors in $M(V)$ (see~(3.1))
that are also the highest weight vectors with respect to the
standard Cartan and Borel subalgebras $\Hw$ and  $\Bw$ of $\gind 0$.

We shall use the following notations:
\[
\La^{\pm}_i:= \La^i(\gind{-1}^{\pm}), \quad
\La^{\pm}:= \sum_{i\geq 0}\La^{\pm}_i\,, \quad
 \Sym^k:=\text{Sym}^k(\gind{-2})\, ,\quad
 \Sym=\sum_{k \geq 0} \Sym^k.\]

We know that $M(V)=U(L_-)\otimes V$ and by the PBW theorem
we have the isomorphisms of vector spaces (where, as before, we
drop the tensor product signs):
\[
M(V)=\Sym\, \La^-\La^+\, V, \qquad  M(V)=\Sym \,\La^+\La^-\, V.
\]
When we use the first isomorphism, we say
that
the \mpo-order (for elements of $\gind{-1}$) is chosen and when the
second, we speak of the \pmo-order.

\bTh\label{T:sl3-weights}
If an $E(3,6)$-module  is degenerate then
the $sl(3)$-highest weight of $V$
is either $(p,0)$ or $(0,q)$.
\eTh

\prf \/
Suppose that the $s\ell (3)$-highest of $V$ is $(p,q)$ and
$pq\neq 0$ and that
the module $M(V)$ is degenerate.  We have to show that this is impossible. Let $w$ be a  non-trivial
 singular vector, which is a $\fg_0$-highest weight vector.

Using the  \mpo \,  order we write
\[
M(V)=\sum_{m,i,j} \Sym^m\La^-_i\La^+_j\,V ,
\]
where the summands are $sl(3)$-modules,
 let $w=\sum w_{m;i,j}$ be the corresponding decomposition
of $w$. Similarly
\[
M(V)=\sum_{m,i,j} \Sym^m\La^+_j\La^-_i\, V,
\]
and $w=\sum \tilde{w}_{m;j,i}$ is the  decomposition for \pmo order.

Let $n$ be the maximum value of $m$ such that there exists
$w_{m;i,j}\neq 0$, and let $n'$ be the similar maximum for
$\tilde{w}_{m;j,i}$.
\bLe If $j\neq 0$ then $w_{n;i,j}= 0$, and if $i\neq 0$ then
$\tilde{w}_{n';j,i}=0$.
\eLe
\prf  \/
Notice that
\begin{equation}
                                          \label{E:filf0}
e'_0\cdot\Sym^m\La^-_i\La^+_j \,V \subset
\Sym^{m-1}\La^-_{i+1}\La^+_{j}\,V +
\Sym^m\La^-_i\La^+_{j-1}\, V \, .
\end{equation}
This follows from the commutation relations
\[
\bAr{lll}
[e'_0 , \dhind1] =0,        &
[e'_0 , \dudind-1]  =0,      &
[e'_0 , \dudind+1]  =+f_2,\\ {}
[e'_0 , \dhind2 ] =0,       &
[e'_0 , \dudind-2]  =0,     &
[e'_0 , \dudind+2]  =-f_{12}, \\ {} 
[e'_0 , \dhind3 ]=-\dudind-3 ,&
[e'_0 , \dudind-3]  =0,    &
[e'_0 , \dudind+3]  =0.%
\eAr
\]
We denote by
$P_{(m;i,j)}$  the projection onto
$\Sym^m\La^-_i\La^+_j V$ (in the  \mpo ~decomposition),
and we see from \eqref{E:filf0} that for any $i \geq 0$ and $j
\geq 1$ we have:
\[
0 = P_{(n;i,j-1)} e'_0 \,w =P_{(n;i,j-1)} e'_0\, w_{n;i,j}\,.
\]
Let us write
$
w_{n;i,j}=\sum \dhind{}^a l_-^I l_+^J w_a^{IJ}
$
where $a,I,J$ are multi-indices and $|a|=n$, \mx{$|I|=i$}, \mx{$|J|=j$}.
We get
\[
P_{(n;i,j-1)} e'_0\, w_{n;i,j}=
\sum \dhind{}^a l_-^I\, (e'_0 \,(l_+^J\, w_a^{IJ}))=0.
\]
So we conclude that for any given $a,I$,
\[ \sum_{|J|=j} e'_0 \,l_+^J\, w_a^{IJ}=0.
\]
Since each $w_{n;i,j}$ is a highest weight vector for $s\ell
(3)$, the coefficient $\sum_J \ell^J_+ w^{IJ}_a$ of
$\hat{\partial}^a \ell^I_-$ of lowest weight in the expression for
$w_{n;i,j}$ is an $s\ell (3)$-highest weight vector.  Hence, by Lemma~3.13,
\mx{$w_{n;i,j}=0$} for \mx{$j>0$}.

In a similar way the commutation relations for $e'_1$ and
Lemma~3.14 imply the second statement of the lemma.\epf

\bLe
  \label{P:w-w}
 \alphaparenlist
\begin{enumerate}
\item 
$w_{n-k;i,j}=0$ for $j>k$ and
$\tilde{w}_{n'-k;j,i}=0$ for $i>k$.
\item 
$n=n'$.
\item 
If $w_{n-k;i,j}\neq 0$ 
\, or \,
$\tilde{w}_{n-k;j,i}\neq 0$ then $i+j=2k$.
\item 
If $w_{n-k;i,j}\neq 0$ then $j\leq k\leq i$, and
if $\tilde{w}_{n-k;j,i}\neq 0$ then $i\leq k \leq j$.
\item 
If $w_{n-k;i,j}\neq 0$ then $i=j=k$.
\item 
 $sl(2)$ acts trivially on $V$.
\end{enumerate}
\eLe
\bCo\label{wform}
$w=w_{n;0,0}+w_{n-1;1,1}+w_{n-2;2,2}+\ldots$ and $w_{n;0,0}\neq 0$.
\eCo

\prf\/ (a) Let us use induction on $k$.
The case $k=0$ follows from Lemma~4.2.
Equation~\eqref{E:filf0} shows that
\[
0 = P_{(n-k;i,j-1)} e'_0\, w =P_{(n-k;i,j-1)} e'_0\,w_{n-k;i,j}
+ P_{(n-k;i,j-1)} e'_0\, w_{n-k+1;i-1,j-1},
\]
but for $j>k$ the last summand is zero by induction.
Now we can apply Lemma~3.13 as we did above and conclude
that  $w_{n-k;i,j}=0$. The other statement follows
in the same way from Lemma~3.14.

To prove (b) let us notice first that
\begin{equation}\label{E:Lachange}
\La^+_j\La^-_i\, V \subset \La^-_i\La^+_j\, V
+\Sym^1\La^-_{i-1}\La^+_{j-1}\, V
+\Sym^2\La^-_{i-2}\La^+_{j-2}\, V +\ldots.%
\end{equation}
We know that if
$\tilde{w}_{(n'-k);j,i}\neq 0$ then $i\leq k$, therefore
from~\eqref{E:Lachange} it follows that
$$
\tilde{w}_{(n'-k);j,i}\in \sum_{s\leq k}
\Sym^{n'-k+s}\La^-_{i-s}\La^+_{j-s}\,V.
$$
As $s\leq k $,  this implies $n\leq n'$, but
the arguments can be reversed so $n=n'$.

For (c) let us notice that
the
$Y$-eigenvalue of $w_{n-k;i,j}$ and of $\tilde{w}_{n-k;i,j}$ is
equal to
$$y_V - {{\frac{1}{3}}}(i+j)-
{
{\frac{2}{3}}}(n-k)$$
where
$y_V$ is defined by
$Y|_V=y_V\,\mx{Id}_V$. But the eigenvalues are all the same whatever
$i,j,k$ so (c) follows.
Now (d) follows immediately from (a-c).

To get (e) let us consider $\PP (w)$ where
\[
\PP=\sum_{m,i\leq j}P_{(m;i,j)}.
\]
If $\tilde{w}_{n-k;j,i}\neq 0$ then  $i\leq k \leq j$ by (d)
and from \eqref{E:Lachange} it follows that
$$
\PP \,\tilde{w}_{n-k;j,i} = \tilde{w}_{n-k;j,i}.
$$
We conclude that
$
\PP\, w= w.
$
But at the same time $w=\sum w_{n-k;i,j}$ and
because  $i>j$ implies
$\PP \,w_{n-k;i,j}=0$,  it follows that  $w_{n-k;i,j}=0$ for $i>j$.
This proves (e).

Corollary~\ref{wform} follows from (e). To establish (f) we need
the following lemma.
\bLe
Let $h\in \CC[x_1,x_2,x_3]$ of degree $n$ and $g=h x_5\dpind4$.
Then $g(\Sym^{n-k}\La^-_{k}\La^+_{k}\,V)=0$ for $k>0$.
\eLe
\begin{proof}

One has to check it for $n=k=1,2,3$, then
the relation
$\,[hx_5\dpind4,\dpind i]=-(\dpind i h)(x_5\dpind4)$
makes it easy to organize induction on $n-k$.
\end{proof}

Now from the lemma it follows that
$f_3 w_{n;0,0}= (x_5\dpind4)w_{n;0,0}=0$.
On the other hand
{$e_3w=0$}, and using the expression for $w$ from Corollary~\ref{wform}
we conclude that $e_3 w_{n;0,0}=0$, but $w_{n;0,0} \neq 0$.

As a result, because $e_3$ and $f_3$ act
trivially on $\gind{-2}$, we conclude that they act trivially on
all coefficients in $w_{n;0,0}$, which are elements from $V$.
But we know that $V$ is isomorphic to
the tensor product of  irreducible
representations of $sl(2)$ and $sl(3)$.
Therefore the existence of
a trivial $sl(2)$ submodule in $V$ means that
$sl(2)$ acts trivially on $V$, which gives (f).\epf \\


Unless otherwise stated, we \emph{use the \mpo-order}.
In the following  we can suppose that $V$ is realized as a submodule in
$\Mw$, i.e.,~that
elements of $V$ are linear combinations
of monomials
\[ [\prod_{i,j} \dpind j^{n_j} x_i^{m_i}],
\]
because the action of $sl(2)$ is trivial due to Lemma~4.3f.  For
$\alpha \in \ZZ^3_+$ we let, as before,
$D^{\alpha}=D^{\alpha_1}_1 D^{\alpha_2}_2 D^{\alpha_3}_3$.

According to Proposition~\ref{P:Dstr-tilde} one has
\begin{equation}                                      \label{E:w-DD}
  w= \sum_{\alpha} D^{\al} T_\al,
\end{equation}
where $T_{\alpha}$ are highest weight vectors in $ \La^-\La^+ V$,
and for their weights we have the relation
\begin{displaymath}                                  
\wt_3 w = (-\al_1+\al_2,\, -\al_2+\al_3)+ \wt_3 T_\al.
\end{displaymath}

If $|\si|=n$ and $T_\si \neq 0$,
then, because of Corollary~\ref{wform},
$T_\si \in V$, so $\wt_3 T_\si = (p,\,q) $, thus
\begin{equation}                                      \label{eq:4.4}
\wt_3 w = (-\si_1+\si_2,\, -\si_2+\si_3)+ (p,\,q).
\end{equation}
This means that given $n$, $\wt_3 w$ and $(p,q)$ we have a unique choice
for $\si$, and we can write  $T_\si=[\dpind3^{q}x_1^p]s$, where
$s$ is a non-zero scalar.  Therefore
\[
 w_n = w_{n;0,0}= D^{\si} [\partial^q_3 x^p_1 ] s, \,\, s \in \CC
 \, .
\]
At the same time, due to (3.3)--(3.5) we have:
\begin{displaymath}
  D^{\sigma} [\partial^q_3 x^p_1]
    = \bar{D}^{\sigma}[\partial^{q-\sigma_2}_3 x^{p-\sigma_1}_1]
      \bar{s} \, ,
\end{displaymath}
where
$\bar{s}=p^{[\si_1]}(p+q+1)^{[\si_1]}q^{[\si_2]}\, s$.



Without loss of generality we can assume that $\bar{s}=1$.  Let
$t_{\sigma} = [\partial^{q-\sigma_2}_3 x^{p-\sigma_1}_1]$.
Using relations (4.2), we compute:
\begin{eqnarray}
  \label{eq:4.5}
  e'_0 \cdot w_n &=& e'_0 \bar{D}^{\sigma} t_{\sigma}
    = - \sigma_1 d^-_3 \bar{D}^{\sigma_1-1}_1
    \bar{D}^{\sigma_2}_2 \bar{D}^{\sigma_3}_3 [x_3] t_{\sigma} \\
\nonumber
  && + \sigma_2 d^-_3 \bar{D}^{\sigma_1}_1 \bar{D}^{\sigma_2-1}_2
     \bar{D}^{\sigma_3}_3 [\partial_2] t_{\sigma}
     -\sigma_3 d^-_3 \bar{D}^{\sigma_1}_1 \bar{D}^{\sigma_2}_2
       \bar{D}^{\sigma_3-1}_3 t_{\sigma} \, .
\end{eqnarray}

Let $P_m = \sum_{i,j} P_{(m;i,j)}$.  It follows from
(\ref{eq:4.5}) that
\begin{displaymath}
  e'_0 \cdot w_k = P_{n-1} e'_0 \cdot w_n \, .
\end{displaymath}
We will use this formula in the following way.  As (4.1) shows,
\begin{displaymath}
 P_{n-1} e'_0 w = P_{n-1} e'_0 w_n + P_{n-1} e'_0 w_{n-1}
\end{displaymath}
and we have $e'_0 w=0$, hence
\starlist

\begin{enumerate}{}{}
  \item \hspace*{1.65in}$  P_{n-1} e'_0 w_n =-P_{n-1} e'_0 w_{n-1}$.
  \end{enumerate}

We already have quite an explicit expression for the left-hand side.
We will write a similar expression for the right-hand side and study
the restrictions imposed by the equality (*).  We will see that
there are very few solutions for these equations in our context
and in the end that no one of them makes it to the
singular highest weight vector.

We know that
\begin{equation}
  \label{eq:4.6}
  w_{n-1} = w_{n-1;1,1}= \sum_{| \beta |=n-1}
  D^{\beta} T_{\beta} \in \Sym \Lambda^-_1 \Lambda^+_1 V \, ,
\end{equation}
where $T_{\beta}$ are the $s\ell (3)$ highest weight vectors in
$\Lambda^-_1 \Lambda^+_1 V $.

\begin{Lemma}
  \label{lem:4.6}
Let $|\beta| =n-1$ and $T_{\beta} \neq 0$.  There are at most six
choices for $\sigma -\beta $: $(-1,1,1)$, $(0,0,1)$, $(0,1,0)$,
$(1,-1,1)$, $(1,0,0)$, $(1,1,-1)$.
\end{Lemma}

\begin{proof}
  It is clear that
  \begin{displaymath}
    \wt_3 w= \wt_{3} w_{n-1} = (-\beta_1+\beta_2, -\beta_2
    +\beta_3) + (\lambda_1,\lambda_2) + (p,q) \, ,
  \end{displaymath}
where $\lambda = (\lambda_1,\lambda_2)$ is a weight of
$\Lambda^-_1 \Lambda^+_1$ and there are six of these weights:
$(2,0)$, $(0,1)$, $(1,-1)$, $(-2,2)$, $(-1,0)$, $(0,-2)$.  But, by
(\ref{eq:4.4}), $\wt_3 w= (-\sigma_1 + \sigma_2 , -\sigma_2 +
\sigma_3) + (p,q)$ as well, so given $\lambda$ we have two linear
equations on $\beta$.  The fact that $|\beta|=n-1$ provides the
third equation and thus the difference $\sigma - \beta$ is
determined.  We get the six values for $\sigma - \beta$ that
correspond to the above six choices for $\lambda$.

\end{proof}

\begin{Lemma}
  \label{lem:4.7}
There are the following possibilities for $T_{\beta}$ (where $t_i
\in \CC$):

\arabicparenlist
\begin{enumerate}
\item 
  $\beta^{(1)} =\sigma - (-1,1,1)$,
     $\wt_3 T_{\beta^{(1)}} = (p,q) + (2,0)$, and
     $T_{\beta^{(1)}} = d^-_1 d^+_1  [\partial^q_3 x^p_1] t_1$.

\vspace{1ex}

\item 
  $\beta^{(2)} =\sigma -(0,0,1) $, $\wt_3 T_{\beta^{(2)}}
     =  (p,q) + (0,1)$, and\hfill\break
   $T_{\beta^{(2)}} =d^-_1 (d^+_1 [x_2]-d^+_3 [x_1])
   [\partial^q_3 x^p_1] t'_2 + (d^-_1 d^+_2 -d^-_2d^+_1)
   [\partial^q_3 x^p_1] t''_2$.

\vspace{1ex}

\item 
  $\beta^{(3)} =\sigma -(0,1,0) $, $\wt_3 T_{\beta^{(3)}}
   =(p,q) + (1,-1)$, and\hfill\break
   $T_{\beta^{(3)}} =d^-_1 \Delta^+ [\partial^{q-1}_3 x^p_1]
   t'_3 + \Delta^- d^+_1 [\partial^{q-1}_3 x^p_1] t''_3$.

\vspace{1ex}

\item 
$  \beta^{(4)} =\sigma - (1,-1,1), \wt_3 T_{\beta^{(4)}}
 =(p,q) + (-2,2)$, and \hfill\break
$T_{\beta^{(4)}} = (d^-_1 [x_2]-d^-_2 [x_1])
 (d^+_1 [x_2]-d^-_2 [x_1]) [\partial^q_3 x^{p-2}_1] t_4$.

\vspace{1ex}

\item 
  $\beta^{(5)} =\sigma - (1,0,0), \wt_3 T_{\beta^{(5)}}
    =(p,q) + (-1,0)$, and \hfill\break
    $T_{\beta^{(5)}} = \left( d^-_1 (d^+_2[x_3] -d^+_3 [x_2] )
      + d^-_2 (d^+_3 [x_1] - d^+_1 [x_3])+\right.$\hfill\break
$\left. \qquad\qquad\qquad\qquad +d^-_3 (d^+_1[x_2] -d^+_2[x_1])\right)
        [\partial^q_3 x^{p-1}_1]t'_5+$\hfill\break
$\left.\qquad\qquad+(d^-_1 [x_2]-d^-_2 [x_1]\right) \Delta^+
   [\partial^{q-1}_3 x^{p-1}_1] t''_5$.  \vspace{1ex}

\item 
  $\beta^{(6)}=\sigma -(1,1,-1), \wt_3 T_{\beta^{(6)}}
    =(p,q) + (0,-2)$, and
    $T_{\beta^{(6)}} =\Delta^- \Delta^+
    [\partial^{q-2}_3 x^p_1] t_6$.
\end{enumerate}
\end{Lemma}

\begin{proof}
  The fact that $T_{\beta}$ is the highest weight vector in
  $\Lambda^-_1 \Lambda^+_1V$ and Lemma~3.12 permit us to write
  $T_{\beta}$ explicitly as soon as its weight is known.  This
  directly leads us to the above expressions.
\end{proof}

Our next step is to look at the lexicographically highest terms
on the left and right hand sides of (*).

\begin{Lemma}
  \label{lem:4.8}
$\ell ht P_{n-1}e'_0 (D^{\beta}T_{\beta})= \hat{\partial}^{\beta} e'_0
h^{[\beta_1]} h^{[\beta_1]}_1 h^{[\beta_2]}_2 T_{\beta}$.
\end{Lemma}

\begin{proof}
  This follows from $| \beta | =n-1$ and Proposition~3.3.
\end{proof}

We can rewrite the lemma as
\begin{displaymath}
  \ell ht P_{n-1} e'_0 (D^{\beta} T_{\beta}) \sim
  \hat{\partial}^{\beta} e'_0 T_{\beta} \, ,
\end{displaymath}
where $\sim$ means equality up to a constant multiple, because as
we know from (3.3)--(3.5), $h^{[\beta_1]}$, $h_1^{[\beta_1]}$,
$h_2^{[\beta_2]}$ multiplies $T_{\beta}$ by a non-zero constant as
long as $T_{\beta} \neq 0$.

Thus we see that if $t_1 \neq 0$, then the $\ell ht$ of the right-hand side of (*) comes from $T_{\beta^{(1)}}$ and is
proportional to
\begin{displaymath}
  \hat{\partial}^{\beta^{(1)}} d^-_1 (-q [\partial_2 \partial_3^{q-1}
  x_1^p] t_1)
\end{displaymath}
(cf. proof of Lemma~3.13).  But the $\ell ht$ of the left-hand
side of (*) is smaller, as one concludes immediately from
(\ref{eq:4.5}) and (3.6).  This implies that $t_1 =0$, and then
the $\ell ht$ on the right side of (*) comes from
$T_{\beta^{(2)}}$ (if it is non-zero), and the $\ell ht$ in
(\ref{eq:4.5}) are clearly the terms with
$\hat{\partial}^{\sigma_1}_1 \hat{\partial}^{\sigma_2}_2
\hat{\partial}^{\sigma_3 -1}_3$.  Comparing the coefficients of this
monomial in (\ref{eq:4.5}) we get:
\begin{eqnarray*}
  \sigma_3 d^-_3 [\partial^q_3 x^p_1] \sim e'_0
     (d^-_1 (d^+_1 [\partial^q_3 x^{p-1}_1 x_2] t'_2
     -d^+_2 [\partial^q_3 x^p_1] t'_2 ) + (d^-_1 d^+_2 - d^-_2
    d^+_1)
    [\partial^q_3 x^p_1] t''_2) \, , \\
\noalign{\hbox{or}}\\
    \sigma_3 d^-_3 [\partial^q_3 x^p_1] \sim - d^-_1
    (f_2 [\partial^q_3 x^{p-1}_1 x_2] t'_2
     + f_{12} [\partial^q_3 x^p_1] t'_2)  +
    (d^-_1 f_{12} + d^-_2 f_2) [\partial^q_3 x^p_1] t''_2 \, .
\end{eqnarray*}
This clearly implies that $\sigma_3 =0$, $t'_2 =t''_2 =0$.

Taking this into account, we can rewrite (*):
\begin{eqnarray}
  \label{eq:4.7}
  \sigma_2 d^-_3 \bar{D}^{\sigma_2 -1}_2
     [\partial_2 \partial^{q-\sigma_2}_3 x^{p-\sigma_1}_1]
   - \sigma_1 d^-_3 \bar{D}^{\sigma_1-1}_1 \bar{D}^{\sigma_2}_2
     [\partial^{q-\sigma_2}_3 x^{p-\sigma_1}_1 x_3]=\\
\nonumber
   = -P_{n-1} e'_0\left(  D^{\sigma_1}_1 D^{\sigma_2-1}_2
       T_{\beta^{(3)}} + D^{\sigma_1-1}_1 D^{\sigma_2}_2
       T_{\beta^{(5)}}
    +D^{\sigma_1-1}_1 D^{\sigma_2-1}_2 D_3
      T_{\beta^{(6)}}\right) \, ,
\end{eqnarray}
where $T_{\beta^{(4)}}$ is absent because there are no such terms
when $\sigma_3 =0$, as the components of $\beta^{(4)}$ are non-negative.

Let us be more careful with the constant factors here.  In
computing the $\ell ht$ on the right side we apply
Lemma~\ref{lem:4.8} to $\beta^{(3)}$ and we get:
\begin{eqnarray*}
  \ell ht P_{n-1} e'_0 (\Dw^{\sigma_1}_1 \Dw^{\sigma_2-1}_2  T_{\beta^{(3)}})
    &=& \hat{\partial}^{\sigma_1}_1 \hat{\partial}^{\sigma_2 -1}_2
      e'_0 h^{[\sigma_1]} h^{[\sigma_1]}_1 h^{[\sigma_2-1]}_2
      T_{\beta^{(3)}}\\
      &=& \hat{\partial}^{\sigma_1}_1 \hat{\partial}^{\sigma_2-1}_2
      e'_0 T_{\beta^{(3)}} (p+1)^{[\sigma_1]}
      (p+q+1)^{[\sigma_1]} (q-1)^{[\sigma_2-1]}
\end{eqnarray*}
because  $\wt_3 T_{\beta^{(3)}} = (p+1,q-1)$
 (and of course $(q-1)^{[\sigma_2-1]} \neq 0$
as $\sigma_2 \leq q)$).
Letting $ b = ((p+1)^{[\sigma_1]} (p+q+1)^{[\sigma_1]}
(q-1)^{[\sigma_2-1]})^{-1}$, we arrive at the following equation:

\begin{eqnarray*}
  \sigma_2 q d^-_3 [\partial_2 \partial^{q-1}_3 x^p_1]
     b &=& d^-_1 e'_0\left( d^+_1 [\partial_1 \partial^{q-1}_3 x^p_1] t'_3
   + d^+_2 [\partial_2 \partial^{q-1}_3 x^p_1] t'_2
       +d^+_3 [\partial^q_3 x^p_1] t'_3 \right)
\\
&&    + d^-_1 e'_0 d^+_1
       [\partial_1 \partial^{q-1}_3 x^p_1]t''_3
%
   +d^-_2 e'_0 d^+_1 [\partial_2 \partial^{q-1}_3 x^p_1] t''_3
     + d^-_3 e'_0 d^+_1 [\partial^q_3 x^p_1] t''_3 \\
  &=& d^-_1 \left( f_2 [\partial_1 \partial^{q-1}_3 x^p_1] t'_3
       -f_{12} [\partial_2 \partial^{q-1}_3 x^p_1] t'_3
     +0 +f_2 [\partial_1 \partial^{q-1}_3 x^p_1] t''_3
     \right)\\
   &&  + d^-_2 f_2 [\partial_2 \partial^{q-1}_3 x^p_1] t''_3
     + d^-_3 f_2 [\partial^q_3 x^p_1] t''_3 \, .
\end{eqnarray*}
Looking at the coefficients of $d^-_1$, we conclude that
$pt'_3=0$ and since $p \neq 0$, $t'_3=0$.  From the coefficients
of $d^-_2$ we see that $(q-1)t''_3 =0$, and from the coefficients
of $d^-_3$ we conclude that $q\sigma_2 b =-qt''_3$.

Thus, either
$\sigma_2 =t'_3 =t''_3=0\,$ or $\,\sigma_2>0$,
   $q=1$, $t'_3=0$, $t''_3 =-\sigma_2
b$. Since $\sigma_2 \leq q$, in the latter case we have
$\sigma_2=1$, $t''_3=-b$.\\

If $\sigma_2 =0$, then (\ref{eq:4.7}) reduces to
\begin{equation}
  \label{eq:4.8}
  \sigma_1d^-_3 \bar{D}^{\sigma_1-1}_1
  [\partial^q_3 x^{p-\sigma_1}_1 x_3] \sim
  P_{n-1} e'_0 D^{\sigma_1-1} T_{\beta^{(5)}} \, .
\end{equation}
Now we look at the coefficients of $\sigma_1^{\sigma_1-1}$ in the
equation.  In the left-hand side we get:
\begin{displaymath}
  \sigma_1 d^-_3 [\partial^q_3 x^{p-1}_1 x_3] \, .
\end{displaymath}
Furthermore, since $\sigma_1=n$, we can use Lemma 4.8
 on the right of (\ref{eq:4.8}), hence the
coefficient of $\hat{\partial}^{\sigma_1-1}_1$ on the right of
(\ref{eq:4.8}) is equal to $e'_0 T_{\beta^{(5)}}$, which we now
compute:
\begin{eqnarray*}
  e'_0 T_{\beta^{(5)}}
      &=& -d^-_1 e'_0 (d^+_2 [\partial^q_3 x^{p-1}_1 x_3]
         -d^+_3 [\partial^q_3 x^{p-1}_1 x_2]) t'_5\\
      && -d^-_2 e'_0 (d^+_3 [\partial^q_3 x^p_1]
         -d^+_1 [\partial^q_3 x^{p-1}_1 x_3]) t'_5\\
      && -d^-_3 e'_0 (d^+_1 [\partial^q_3 x^{p-1}_1 x_2]
         -d^+_2 [\partial^q_3 x^p_1]) t'_5 \\
      && -d^-_1 e'_0 \Delta^+ [\partial^q_3 x^{p-1}_1 x_2] t''_5
        -d^-_2 e'_0 \Delta^+ [\partial^{q-1}_3 x^p_1] t''_5\\
      &=& d^-_1 f_{12} [\partial^q_3 x^{p-1}_1 x_3] t'_5
        + d^-_2 f_2 [\partial^q_3 x^{p-1}_1 x_3] t'_5 \\
      &&  -d^-_3 (f_2 [\partial^q_3 x^{p-1}_1 x_2] t'_5
           + f_{12} [\partial^q_3 x^p_1] t'_5) \\
      &&  -d^-_1 (f_2 [\partial_1 \partial^{q-1}_3 x^{p-1}_1x_2]
          t''_5 - f_{12} [\partial_2 \partial^{q-1}_3 x^{p-1}_1
          x_2] t''_5 )\\
      && -d^-_2 (f_2 [\partial_1 \partial^{q-1}_3 x^p_1] t''_5
         -f_{12} [\partial_2 \partial^{q-1}_3 x^p_1] t''_5 ) \, .
\end{eqnarray*}
At the end we get for $e'_0 T_{\beta^{(5)}}$:
\begin{eqnarray}
  \nonumber
  \hspace*{3ex} e'_0 T_{\beta^{(5)}}
     &=&\, d^-_1 \left(\, (-qt'_5-t''_5) [\partial_1x_1]^{} + (p-1) t''_5
         [\partial_2x_2]+(p-1)t'_5 [\partial_3x_3]^{}\, \right)
         [\partial^{q-1}_3 x^{p-2}_1 x_3]\\
  \label{eq:4.9}
     && +\, d^-_2 (-qt'_5 -pt''_5)
         [\partial_2 \partial^{q-1}_3 x^{p-1}_1x_3]
      + d^-_3 ((p+q+1)t'_5) [\partial^q_3 x^{p-1}_1 x_3] \, .
\end{eqnarray}
We conclude that the terms with $d^-_1$ and $d^-_2$ disappear iff
either $p=1, qt'_5 +t''_5 =0$ or when $T_{\beta^{(5)}}=0$.  Now
for the case when $\sigma_3=\sigma_2=0$ and $T_{\beta^{(5)}}=0$
we get $\sigma_1=0$ and this is a contradiction.

If $\sigma_2=\sigma_3=0$ and $p=1$, then $\sigma_1=1$ because
$\sigma_1 \leq p$ and it could not be $0$ as this gives $|\sigma
|=0$.  So it becomes $|\sigma|=n=1$ and $w=w_n+w_{n-1}=w_1+w_0
=D_1 [\partial^q_3] +T_{\beta^{(5)}}$ ($T_{\beta^{(6)}}$
disappears for $\sigma_2=0$).  We can check $e_3 \cdot w$ now:
\begin{eqnarray*}
  e_3 \cdot w &=& 0+e_3 T_{\beta^{(5)}}\\
   &=& 2  (d^+_{12}[x_3] +d^+_{23} [x_1] +d^+_{31} [x_2])[\partial^q_3]t'_5
   + (d^+_1 [x_2] -d^+_2 [x_1]) \Delta^+
       [\partial^{q-1}_3] t''_5 \, ,
\end{eqnarray*}
therefore $e_3 \cdot w =0$ implies $t'_5=t''_5=0$ and we arrive
at a contradiction.

If $\sigma_2=1$, we are left with the situation when $q=1,
\sigma_2=1, t'_3=0, t''_3=-b$.  Then (\ref{eq:4.7}) reduces to:
\begin{eqnarray}
  \label{eq:4.10}
  d^-_3 \bar{D}^{\sigma_1}_1 [\partial_2 x^{p-\sigma_1}_1]
     -\sigma_1 d^-_3 \bar{D}^{\sigma_1-1}_1 D_2
     [x^{p-\sigma_1}_1 x_3]=\\
\nonumber
      =-P_{n-1} e'_0 (D^{\sigma_1}_1 T_{\beta^{(3)}} +
          D^{\sigma_1-1} D_2 T_{\beta^{(5)}}) \, .
\end{eqnarray}
Note that $\sigma_1\neq0$ because
   $\sigma_1 +1=n$, and, if $\sigma_1=0$, then $n=1$,
the term with $T_{\beta^{(5)}}$ disappears and $w=w_n+w_{n-1}$,
hence we have:
\begin{displaymath}
  w=\bar{D}_2 t_{\sigma}-D_1 \Delta^- d^+_1 [x^p_1] b \, .
\end{displaymath}
Since $e_3 w=0$ and $e_3$ annihilates the first summands but does
not annihilate the second one, we arrive at a contradiction. Therefore
$\sigma_1 \neq 0$.

We already know that the $\ell ht$ in both
sides of (\ref{eq:4.10}) are equal, so let us look at the next
ones, i.e.,~the coefficients of $\hat{\partial}^{\sigma_1-1}_1
\hat{\partial}_2$.  In the left-hand side we get:
\begin{equation}
  \label{eq:4.11}
  \sigma_1 d^-_3 [x^{p-1}_1] ([\partial_2 x_2] -[\partial_3x_3])
  \, .
\end{equation}
In order to do the same for the right-hand side, we need the
second lexicographically ordered term of $D^{\sigma_1}_1$.  Using
Lemma~3.8 twice, we have
\begin{eqnarray*}
  D^{\sigma_1}_1
  &=& A^{\sigma_1} h^{[\sigma_1]} + \cdots
      = \hat{\partial}^{\sigma_1}_1 h^{[\sigma_1]}_1 h^{[\sigma_1]}
      +\sigma_1 \hat{\partial}_2 \hat{\partial}^{\sigma_1-1}_1
         f_1 (h_1-1)^{[\sigma_1-1]} h^{[\sigma_1]} + \ldots \\
  &=& \hat{\partial}^{\sigma_1}_1 h^{[\sigma_1]} +\sigma_1
      \hat{\partial}^{\sigma_1-1}_1 \hat{\partial}_2 (h_1+1)^{[\sigma_1-1]}
      (h+1)^{[\sigma_1]} f_1 + \cdots \, .
\end{eqnarray*}
Hence the coefficient of $\hat{\partial}^{\sigma_1-1}_1 \hat{\partial}_2$
 on the right-hand side of (\ref{eq:4.10}) is:
 \begin{eqnarray*}
   -e'_0 \left((h_1+1)^{[\sigma_1-1]} (h+1)^{[\sigma_1]}
        f_1 T_{\beta^{(3)}}
      + h^{[\sigma_1-1]} h_1^{[\sigma_1-1]} h_2
          T_{\beta^{(5)}}  \right)\, .
 \end{eqnarray*}
Since the weights of $f_1 T_{\beta^{(3)}}$ and  $T_{\beta^{(5)}}$
are both equal to $(p-1,1)$, this becomes:
\begin{equation}
  \label{eq:4.12}
  -(p(p+2) e'_0 f_1 T_{\beta^{(3)}} + (p-\sigma_1 +1)
   e'_0 T_{\beta^{(5)}}) (p-1)^{[\sigma_1-1]}(p+1)^{[\sigma_1-1]}
   \, .
\end{equation}
Comparing (\ref{eq:4.11}) and (\ref{eq:4.12}) gives:
\begin{eqnarray}
  \label{eq:4.13}
   &\sigma_1 d^-_3 [x^{p-1}_1] ([\partial_2 x_2]
    -[\partial_3 x_3])=\\
\nonumber
   &=- (p-1)^{[\sigma_1-1]}
    (p+1)^{[\sigma_1-1]}
    \left(p(p+2) e'_0 f_1 T_{\beta^{(3)}} + (p-\sigma_1+1)
    e'_0 T_{\beta^{(5)}}\right) \, .
\end{eqnarray}
Since $e'_0$ commutes with $f_1$, we can use our previous
calculation of $e'_0 T_{\beta^{(3)}}$:
\begin{displaymath}
  e'_0 T_{\beta^{(3)}} =-d^-_3 [\partial_2 x^p_1]b \, .
\end{displaymath}
Hence
\begin{equation}
  \label{eq:4.14}
   e'_0 f_1 T_{\beta^{(3)}} =
  f_1 e'_0 T_{\beta^{(3)}} =
   d^-_3
  [\partial_1x^p_1] b-p\,d^-_3 [\partial_2 x_2x^p_1] b\, .
\end{equation}
Also, our previous calculation of $e'_0 T_{\beta^{(5)}} $shows
that in general terms with $d^-_1, d^-_2$ are present in $e'_0
T_{\beta^{(5)}}$.  But (\ref{eq:4.13}) shows that these terms
have to be zero because there are no such terms in the other
entries in (\ref{eq:4.13}).  We conclude that $p=1,\, t'_5 + t''_5
=0$.

So we are left with $p=1$, $q=1$, $\sigma_3=0$, $\sigma_2=1$ and
$\sigma_1 \neq 0$.  The latter condition implies $\sigma_1=1$
because $\sigma_1 \leq p$.  Let us write again the coefficients
of $\hat{\partial}^{\sigma_1-1}_1 \hat{\partial}_2 = \hat{\partial}_2$
   in (\ref{eq:4.10}), which are given by (\ref{eq:4.12}),
for our specific data:
\begin{displaymath}
  d^-_3 ([\partial_2x_2]-[\partial_3x_3]) =2 e'_0
     f_1 T_{\beta^{(3)}} -e'_0  T_{\beta^{(5)}} \, .
\end{displaymath}
Together with (\ref{eq:4.14}) and (\ref{eq:4.9}) we arrive at
\begin{displaymath}
  d^-_3 ([\partial_2x_2] -[\partial_3 x_3])
  = d^-_3 (3 ([\partial_1x_1]-[\partial_2x_2]) (-1/6)
  -3  t'_5 [\partial_3 x_3]) \,
\end{displaymath}
%
%
%
which gives $t'_5=1/2$.

Let us calculate also the terms with $\hat{\partial}_3$ at both sides of
(\ref{eq:4.10}).  We get
\begin{eqnarray*}
  \hat{\partial}_3 d^-_3 [\partial_2 x_2]\,2
  &=& -P_1 (e'_0 (\hat{\partial}_3 (f_{12} h_1 +f_2f_1)
       T_{\beta^{(3)}} +\hat{\partial}_3 f_2 T_{\beta^{(5)}}))\\
  &=&   -\hat{\partial}_3 ((f_{12}h_1+f_2f_1) e'_0
       T_{\beta^{(3)}} +f_2 e'_0 T_{\beta^{(5)}})\\
  &=&  +\hat{\partial}_3 ((f_{12}h_1+f_2f_1)
    (d^-_3 [\partial_2x_1]\frac{1}{6}) -f_2
    (d^-_3 [\partial_3x_3] \frac{1}{2})) \, .
\end{eqnarray*}
Clearly
\begin{displaymath}
  (f_{12}h_1+f_2f_1) d^-_3 [\partial_2x_1]
  = d^-_3 (f_{12}h_1+f_2f_1) [\partial_2x_1]
  = d^-_3 [\partial_2x_3]\,3
\end{displaymath}
and
\begin{displaymath}
  f_2 \, d^-_3 [\partial_3 x_3] = d^-_3 \, f_2
  [\partial_3x_3] =-d^-_3 [\partial_2x_3] \, .
\end{displaymath}

Combining these equations, we get:
\begin{displaymath}
  d^-_3 [\partial_3x_3]\, 2 =
  d^-_3 [\partial_2x_3]  \frac{1}{2}
  + d^-_3 [\partial_2x_3]  \frac{1}{2} \, ,
\end{displaymath}
a contradiction.  This closes the last case and ends the proof of
Theorem~4.1.

\vspace{6ex}


\vspace{6ex}

\textbf{Authors' addresses:}
\begin{list}{}{}

\item  Department of Mathematics, MIT,
Cambridge MA 02139,
USA,\\
email:~~kac@math.mit.edu

\vspace{1ex}

\item   Department of Mathematics, NTNU, Gl\o shaugen,
N-7491 Trondheim,
Norway, \\
email:~~rudakov@math.ntnu.no

\end{list}

\end{document}